\documentclass[preprint,onecolumn,nofootinbib]{revtex4}
\pdfoutput=1
\usepackage[colorlinks=true,linkcolor=blue,urlcolor=blue,filecolor=black,citecolor=red,pdfstartview=FitV,pdftitle={},pdfsubject={},pdfkeywords={},pdfpagemode=None,bookmarksopen=true]{hyperref}
\usepackage{graphicx}
\usepackage{amsmath}
\usepackage{amsfonts}
\usepackage{amssymb,ulem}
\usepackage{color,xcolor}%
\usepackage{CJK}
\usepackage{subfigure}

\usepackage{dcolumn}
\newcommand{\f}{\begin{equation}}
\newcommand{\ff}{\end{equation}}
\newcommand{\fa}{\begin{eqnarray}}
\newcommand{\ffa}{\end{eqnarray}}

\begin{document}
\title{Holographic superfluid with gauge-axion coupling}
\author{Yan Liu$^{1}$}
\author{Xi-Jing Wang$^{2}$}
\author{Jian-Pin Wu$^{2}$}
\thanks{jianpinwu@yzu.edu.cn, corresponding author}
\author{Xin Zhang$^{1,3,4}$}
\thanks{zhangxin@mail.neu.edu.cn}
\affiliation{
    $^1$ Key Laboratory of Cosmology and Astrophysics (Liaoning Province) \& Department of Physics, College of Sciences, Northeastern University, Shenyang 110819, China \\
    $^2$ Center for Gravitation and Cosmology, College of Physical Science and Technology, Yangzhou University, Yangzhou 225009, China \\
    $^3$ National Frontiers Science Center for Industrial Intelligence and Systems Optimization, Northeastern University, Shenyang 110819, China\\
    $^4$ Key Laboratory of Data Analytics and Optimization for Smart Industry (Northeastern University), Ministry of Education, Shenyang 110819, China
}
\begin{abstract}
We have constructed a holographic superfluid with gauge-axion coupling. Depending on whether the coupling is positive or negative, the system displays metallic or insulating behavior in its normal state. A significant feature of the system is the appearance of a mid-IR peak in the alternating current (AC) conductivity in a certain range of parameters. This peak arises due to competition between explicit symmetry breaking (ESB) and spontaneous symmetry breaking (SSB), which results in the presence of a pseudo-Goldstone mode. Moreover, a dip in low-frequency AC conductivity is observed, stemming from the excitation of the SSB Goldstone mode. In the superfluid phase, the effect of gauge-axion coupling on the condensation or superfluid energy gap is only amplified in the presence of strong momentum dissipation. Notably, for the case with negative gauge-axion coupling, a hard-gap-like behavior at low frequency and a pronounced peak at intermediate frequency are observed, indicating that the evolution of the superfluid component is distinct from that of positive coupling.

\end{abstract}

\maketitle
\tableofcontents

\section{Introduction}

The AdS/CFT correspondence serves as a link between the weakly coupled classical gravitational theory and strongly coupled quantum field theory (QFT) \cite{Maldacena:1997re,Gubser:1998bc,Witten:1998qj,Aharony:1999ti}. This correspondence has revealed some universal properties in strongly coupled quantum many-body systems and has provided important insights into phenomena such as transport properties without quasiparticle excitations, novel mechanisms for superconductivity, and quantum phase transitions (QPTs). Such achievements represent significant landmarks in the field.

Transport properties are important characteristics of strongly coupled quantum many-body systems. In the holographic framework, extensive studies have been conducted on transport properties, as reviewed in \cite{Hartnoll:2009sz,Natsuume:2014sfa,Hartnoll:2016apf,Baggioli:2019rrs,Baggioli:2021xuv} and elsewhere. Notably, the introduction of momentum dissipation as a mechanism has led to significant progress in modeling more realistic systems. This mechanism removes the $\delta$-function in the electric conductivity at zero frequency in the holographic system without momentum dissipation. It has been applied to study the behavior of strange metals \cite{Hartnoll:2009ns,Davison:2013txa,Anderson,coleman1996should,Zhou:2015dha}, the mechanism for coherent and incoherent metals \cite{Davison:2015bea,Zhou:2015qui}, and the implementation of QPTs \cite{Mott:1968nwb,Ling:2016dck,Nishioka:2009zj}.

Spatially linear dependent scalar fields, known as axionic fields, provide a simple yet significant mechanism for momentum dissipation \cite{Andrade:2013gsa}. In this holographic framework, when the momentum dissipation is weak, a standard Drude metal can be implemented using this axion model. As the momentum dissipation increases, the system transitions to an incoherent phase \cite{Davison:2015bea,Zhou:2015qui}. Notably, when we measure in terms of the chemical potential of the dual field theory, i.e., the system is in the grand canonical ensemble, the direct current (DC) conductivity of the simplest $4$-dimensional holographic axions model without higher-derivative terms, as studied in \cite{Andrade:2013gsa}, is a non-vanishing constant, independent of temperature. Moreover, there exists a lower bound for the DC conductivity in this holographic dual system \cite{Grozdanov:2015qia}. Hence, the metal-insulator transition (MIT) is absent in the holographic axion model.

In the spirit of effective holographic low energy theories, it is natural and intriguing to investigate the impact of higher-derivative terms of axion fields \cite{Gouteraux:2016wxj,Baggioli:2016oqk,Baggioli:2016pia,Li:2018vrz,Huh:2021ppg}. Recent studies have revealed that the holographic effective theory with higher-derivative terms can break the lower bound of DC conductivity in the usual axion model \cite{Andrade:2013gsa}. This provides a framework to model insulating states with vanishing DC conductivity at zero temperature, which is more consistent with realistic systems.
Furthermore, the higher-derivative terms have a significant effect on the lower bound of charge diffusion \cite{Baggioli:2016pia}, while leaving its upper bound and the bound of energy diffusion unaffected \cite{Huh:2021ppg}.

Higher-derivative axion models can be classified into two important classes \cite{Gouteraux:2016wxj,Baggioli:2016pia}:
\begin{itemize}
	\item \textbf{$\mathcal{J}$ model:}
	\fa
	\label{action-J}
	S_{\mathcal{J}}=\int d^4x\sqrt{-g}\Big(R+6-X-\frac{1}{4}F^{2}-\frac{\mathcal{J}}{4}{\rm{Tr}}[XF^{2}]\Big)\,.
	\ffa
	It is found that the $\mathcal{J}$ coupling has no effect on the background such that one has an analytical background, which is just the Reissner-Nordstr{\"o}m-AdS (RN-AdS) black hole solution with axions \cite{Andrade:2013gsa}. But it enters into the perturbative equations and significantly influences the transport properties of this model.
	\item \textbf{$\mathcal{K}$ model:}
	\fa
	\label{action-K}
	S_{\mathcal{K}}=\int d^4x\sqrt{-g}\Big(R+6-X-\frac{1}{4}F^{2}-\frac{\mathcal{K}}{4}XF^{2}\Big)\,.
	\ffa
	Compared to the $\mathcal{J}$ model, the $\mathcal{K}$ coupling term affects both the background solution and the perturbative equations. However, an analytical black hole solution still exists for this theory \cite{Gouteraux:2016wxj}.
\end{itemize}
In both actions mentioned above, the gauge field $A_{\mu}$ is associated with the field strength $F_{\mu\nu}=\nabla_{\mu}A_{\nu}-\nabla_{\nu}A_{\mu}$ in both actions mentioned above.  Additionally, we define $X\equiv {\rm{Tr}}[X^{\mu}\,_{\nu}]$, where 
\fa
X^{\mu}\,_{\nu}=\frac{1}{2}\partial^{\mu}\phi^{I}\partial_{\nu}\phi^{I}\,,
\ffa
with $I$ taking the values of $x$ and $y$. Here, $\phi^I$ represents a pair of spatial linear axionic fields, specifically $\phi^{I}=\alpha\delta^I_i x^i$ ($i=1,2$).
$\mathcal{J}$ and $\mathcal{K}$ are the coupling constants, respectively.

Further studies manifest that in the aforementioned models, both the explicit symmetry breaking (ESB) and the spontaneous symmetry breaking (SSB) can emerge simultaneously \cite{Baggioli:2014roa,Alberte:2017cch,Ammon:2019wci,Li:2018vrz,Wang:2021jfu,Zhong:2022mok}. The SSB is associated with the gapless excitations known as Goldstone modes in the low energy description. In cases where the dominance of SSB surpasses that of ESB, a novel mode referred to as a pseudo-Goldstone mode emerges. For the pseudo-Goldstone physics in holography, please refer to the recent review \cite{Baggioli:2022pyb}, provides a thorough summary of the significant advancements made in this field over the last decade.

Then, the holographic superfluid model with $\mathcal{J}$ coupling is constructed and its superfluid properties are explored \cite{Liu:2022bam}. In the superfluid phase, the $\mathcal{J}$ coupling  plays a key role, leading to a more pronounced gap in the low frequency conductivity. Additionally, we extensively explore the combined effects of broken translations and various couplings among the gauge field, axion fields, and the complex scalar field. In this paper, we shall study the properties of electric conductivity of $\mathcal{K}$ model. Based on this, we construct a holographic superfluid model with $\mathcal{K}$ coupling and further explore its superfluid properties.

The structure of this work is outlined as follows. In Section \ref{sec-setup}, we construct a holographic model that incorporates higher derivative terms of axion fields to enable momentum relaxation. The numerical calculations of the DC and frequency-dependent conductivity (alternating current conductivity, AC conductivity) in the normal state, along with an analysis of the gauge-axion coupling's impact, are presented in Section \ref{Normal}. Section \ref{super} is dedicated to exploring the influence of the higher derivative terms on the condensation and the conductivity of the superfluid phase. Finally, we provide a summary and discuss the obtained results in Section \ref{conclusion}, concluding the paper.

\section{Holographic framework}\label{sec-setup}

We propose an effective holographic superfluid model with a generalized $\mathcal{K}$ coupling, whose action is given by
\fa
&&
\label{action}
S=\int d^4x\sqrt{-g}\Big(R+6+\mathcal{L}_{M}+\mathcal{L}_{\chi}+\mathcal{L}_{X}\Big)\,.
\ffa
Here, $\mathcal{L}{M}$, $\mathcal{L}{\chi}$, and $\mathcal{L}_{X}$ represent the Lagrangian densities for the Maxwell field, the charged complex scalar field that supports the superfluid phase, and the axionic fields, respectively:
\fa
&&
\label{L}
\mathcal{L}_{M}=-\frac{Z(\chi)}{4}F^{2}\,,
\
\\
&&
\mathcal{L}_{\chi}=-\frac{1}{2}(\partial_{\mu} \chi)^{2}
-H(\chi)(\partial_{\mu}\theta-qA_{\mu})^{2}-\frac{M^{2}\chi^{2}}{2}\,,
\
\\
&&
\mathcal{L}_{X}=-\frac{1}{4}\mathcal{K}(\chi)XF^{2}-X\,.
\ffa
In the above action, we decompose the charged scalar field $\psi$ into a real scalar field $\chi$ and a St\"uckelberg field $\theta$, represented as $\psi = \chi e^{i\theta}$. The parameter $M$ corresponds to the mass of the scalar field $\psi$. For simplicity, we adopt the gauge $\theta=0$ in the subsequent analysis.

We assume that the coupling functions in the action take the form as: $Z(\chi)=1+\beta\chi^{2}/2$, $H(\chi)=n\chi^{2}/2$ and $\mathcal{K}(\chi)=\mathcal{K}_{1}+\mathcal{K}_{2}\chi^{2}/2$. Here, $\beta$, $n$, $\mathcal{K}_1$, and $\mathcal{K}_2$ denote the corresponding parameters for the higher derivative couplings. It is important to note that $\mathcal{K}(\chi)$ represents the generalization of the $\mathcal{K}$ coupling in the $\mathcal{K}$ model (Eq.~\eqref{action-K}).

From the action \eqref{action}, we obtain the equations of motion as follows:
\fa
\label{Maxwell}
&&
\nabla_{\mu}\Big[(Z+\mathcal{K}X)F^{\mu\nu}\Big]-2Hq^{2}A^{\mu}=0\,,
\\
&&
\nabla_{\mu}\nabla^{\mu}\chi-\partial_{\chi}Hq^{2}A^2-\frac{\partial_{\chi}Z}{4}F^{2}-\frac{\partial_{\chi}\mathcal{K}}{4}XF^{2}-M^{2}\chi=0,
\\
&&
\nabla_{\mu}[\nabla^{\mu}\phi^{I}-\frac{\mathcal{K}}{4}(\nabla^{\mu}\phi^{I})F^{2}]=0\,,
\ffa
and
\fa
\label{Einstein}
&&
R_{\mu\nu}-\frac{1}{2}g_{\mu\nu}R-3g_{\mu\nu}-\frac{1}{2}\nabla_{\mu}\chi\nabla_{\nu}\chi-Hq^{2}A_{\mu}A_{\nu}-\frac{1}{2}(Z+\mathcal{K}X)F_{\mu\rho}F_{\nu}\,^{\rho}
\nonumber
\\
&&
-\frac{1}{2}(1+\frac{\mathcal{K}}{4}F^2)\nabla_{\mu}\phi^{I}\nabla_{\nu}\phi^{I}+\frac{1}{8}(Z+\mathcal{K}X)g_{\mu\nu}F^2+\frac{1}{2}g_{\mu\nu}X
\nonumber
\\
&&
+\frac{1}{4}g_{\mu\nu}\nabla_{\mu}\chi\nabla^{\mu}\chi+\frac{1}{2}g_{\mu\nu}Hq^{2}A^{2}+\frac{1}{4}g_{\mu\nu}M^{2}\chi^{2}=0\,.
\ffa

To numerically solve the above equations, it is advantageous to adopt the following ansatz:
\fa
&&
\label{black hole background solution}
ds^2={1\over u^2}\left[-(1-u)p(u)U_{1}dt^2+\frac{du^2}{(1-z)p(u)U_{1}}+U_{2}dx^2+U_{2}dy^2\right],
\\
&&
A_{t}=\mu(1-u)a(u)\,,\,\,\,\,\chi=u^{3-\Delta}\phi\,,\,\,\,\,
\\
&&
\label{special solution}
\phi^{x}=\alpha x\,,\,\,\,\,\phi^{y}=\alpha y\,,
\ffa
where $p(u)=1+u+u^2-\mu^2u^3/4$ and $\Delta=3/2\pm(9/4+M^2)^{1/2}$. Throughout this paper, we set $M^2=-2$ to ensure compliance with the ${\rm AdS}{2}$ Breitenlohner-Freedman (BF) bound. Notably, the functions $U{1}$ and $U_{2}$ solely depend on the radial coordinate $u$. Additionally, we interpret $\mu$ as the chemical potential and $\alpha$ as the strength of momentum dissipation. By imposing the boundary condition $U(1)=1$ at the horizon, the Hawking temperature of the system can be expressed as follows:
\fa
\label{tem}
T=\frac{3}{4\pi}-\frac{\mu^2}{16\pi}\,.
\ffa
For given coupling parameters, the system can be characterized by the dimensionless quantities $T/\mu$ and $\alpha/\mu$. To simplify notation, we will continue using the shorthand ${T,\alpha}$ to represent the dimensionless quantities ${T/\mu,\alpha/\mu}$ throughout the rest of the paper.

\section{Conductivity in normal phase}\label{Normal}

The properties of DC conductivity in the normal phase have been extensively explored in \cite{Gouteraux:2016wxj}. However, the investigation of AC conductivity in this context is still lacking. Therefore, in this section, our primary focus will be on studying the properties of AC conductivity in the normal phase. To provide a comprehensive analysis, we will also provide a concise review of DC conductivity.

Before proceeding, we will analyze the asymptotic behavior of the scalar field $\phi^{I}$ near the AdS boundary to explain the role of the two terms in the Lagrangian density $\mathcal{L}_{X}$. This analysis closely follows similar approaches in previous works such as \cite{Alberte:2017oqx, Li:2018vrz, Wang:2021jfu, Zhong:2022mok, Amoretti:2016bxs, Amoretti:2017frz, Amoretti:2017axe, Amoretti:2018tzw, Amoretti:2019cef}. When we focus solely on the second term, $X$, in $\mathcal{L}_{X}$, the asymptotic expansion of $\phi^{I}$ at the UV boundary can be described as \cite{Wang:2021jfu}
\fa
\phi^{I}=\phi_{(0)}^{I}(t,x^{i})+\phi_{(3)}^{I}(t,x^{i}) u^{3}+\cdots\,.
\ffa
In accordance with standard quantization, the leading order term is identified as the external source for the dual scalar operator $\mathcal{O}^{I}$ and follows the behavior of the special solution (\ref{special solution}), specifically $\phi_{(0)}^{I}(t,x^{i})=\alpha x^{I}$. It plays the role of ESB, responsible for momentum relaxation on the field theory side.

When only the first term $\mathcal{K}XF^{2}$ of $\mathcal{L}_{X}$ survives, at the UV boundary, the scalar field $\phi^{I}$ can be expressed as \cite{Wang:2021jfu}:
\fa
\phi^{I}=\phi^{I}_{(-1)}(t,x^{i})u^{-1}+\phi^{I}_{(0)}(t,x^{i})+\cdots\,.
\ffa
In this case, the subleading order $\phi^{I}{(0)}(t,x^{i})$ corresponds to the non-zero expectation value $\langle\mathcal{O}^{I}\rangle$, where the source term vanishes~\cite{Li:2018vrz}. It follows the behavior of the special solution (\ref{special solution}), i.e., $\langle\mathcal{O}^{I}\rangle\sim\phi_{(0)}^{I}(t,x^{i})=\alpha x^{I}$. This term plays the role of the spontaneous breaking of translations, leading to the emergence of gapless excitations known as Goldstone modes in the low energy description.

By manipulating the couplings in this model, we can readily achieve diverse forms of symmetry breaking. When the first term in $\mathcal{L}_{X}$ dominates over the second term, translations are explicitly but weakly broken in the system. In other words, the breaking of translations is predominantly explicitly but not entirely. In the field theory perspective, when a spontaneously broken symmetry is only approximate, a slight gap emerges in the system, leading to the presence of a pseudo-Goldstone mode. It is worth noting that $\mathcal{K}_{1}$ controls the
SSB, while $\alpha$ controls both the ESB and SSB in our setup.

Next, we will study the DC and AC conductivities on the normal state, respectively.

\subsection{DC conductivity}

We can utilize the widely employed ``membrane paradigm'' \cite{Iqbal:2008by,Donos:2014cya,Amoretti:2014mma,Donos:2015gia} to analytically derive the expression for the DC conductivity, which relies on the data of the IR geometry. The central idea of this approach involves establishing a radially conserved current that connects the data at the horizon and the boundary. Based on this point, we can calculate the DC conductivity of the corresponding dual boundary system by leveraging the horizon data, yielding the following result:
\fa
\label{DCconductivity}
\sigma_{DC}=-\frac{(\alpha^2\mu^{2}\mathcal{K}_{1}+U_{2}(1))(\alpha^2\mu^{2}(2+\mathcal{K}_{1}\mu^{2}a^{2}(1))+2\mu^{2}a^{2}(1)U_{2}(1))}{\alpha^2\mu^{2}(-2+\mathcal{K}_{1}\mu^{2}a^{2}(1))U_{2}(1)}.
\ffa
Equipped with equation \eqref{DCconductivity}, we investigate the impact of higher-derivative axion terms on the DC conductivity. It is worth noting that the gauge-axion coupling must satisfy the constraints $-1/6\leq\mathcal{K}_{1}\leq1/6$, as dictated by the null energy condition (NEC) and particle-vortex duality \cite{Gouteraux:2016wxj,Baggioli:2016pia}.
\begin{figure}
\center{
\includegraphics[scale=0.65]{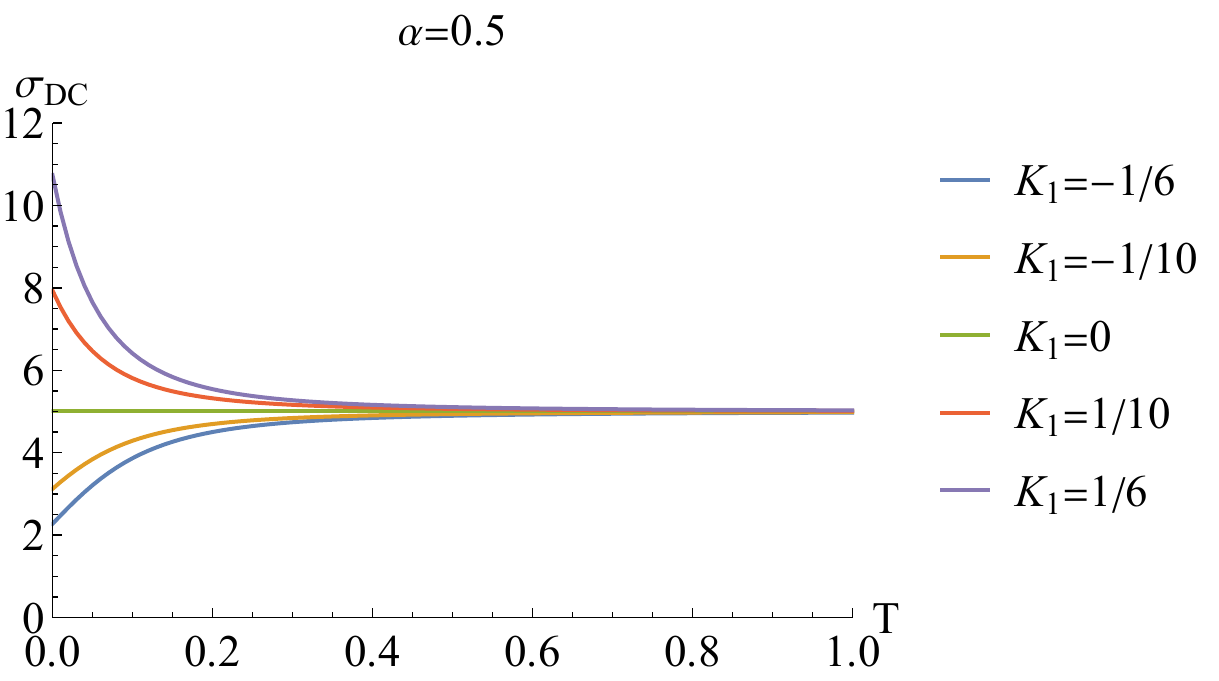}\hspace{0.1cm}
\includegraphics[scale=0.65]{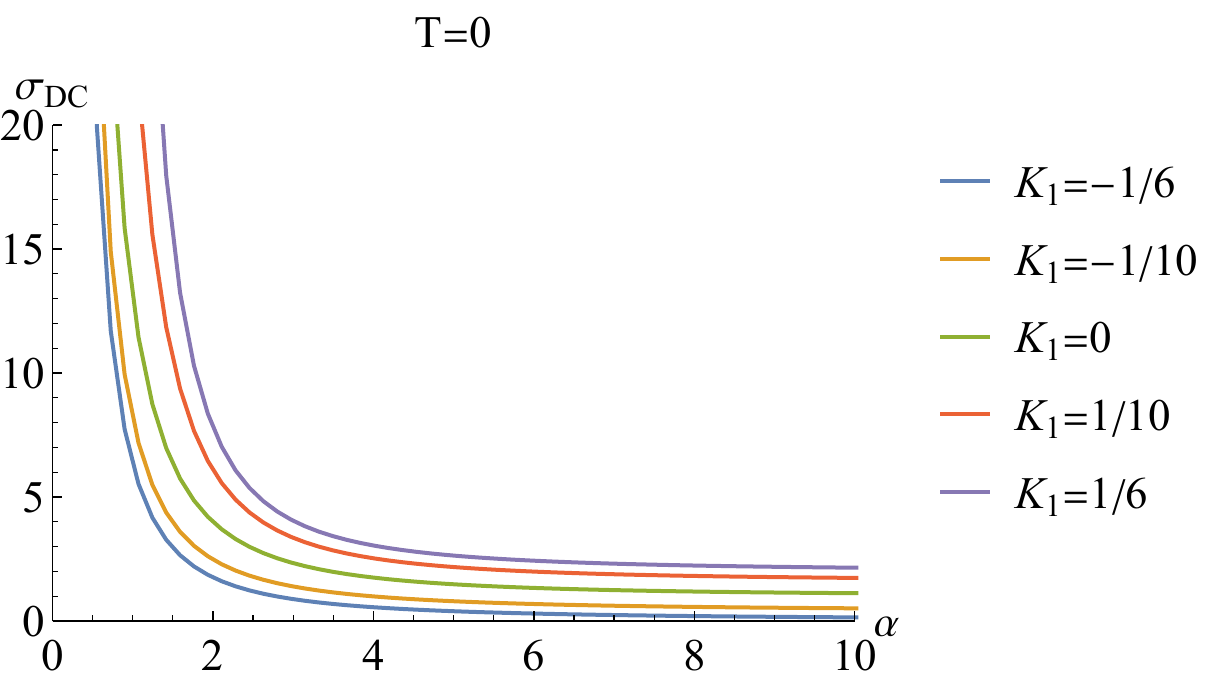}
\caption{\label{DCcalculate} The behaviors of DC conductivity for various coupling parameters. The left plot showcases the temperature behaviors of the DC conductivity for different gauge-axion coupling parameters $\mathcal{K}{1}$, while keeping the momentum dissipation strength $\alpha=0.5$. The right plot displays the DC conductivity as a function of $\alpha$ for different $\mathcal{K}_{1}$ values at a fixed temperature of $T=0$.
}}
\end{figure}

In the left plot of Fig.~\ref{DCcalculate}, we present the temperature dependencies of the DC conductivity for various gauge-axion coupling parameters, while maintaining a constant momentum dissipation strength. In the case where the gauge-axion coupling parameter $\mathcal{K}_{1}=0$, the system simplifies to the axions model \cite{Andrade:2013gsa}, and the DC conductivity becomes temperature-independent. However, when we introduce a non-zero gauge-axion coupling ($\mathcal{K}_{1}\neq0$), the DC conductivity becomes temperature-dependent, as observed in \cite{Gouteraux:2016wxj}. Specifically, for $\mathcal{K}_{1}>0$, the DC conductivity increases with decreasing temperature, indicating a metallic behavior in the dual system. Conversely, for $\mathcal{K}_{1}<0$, the DC conductivity exhibits the opposite temperature dependence, suggesting an insulating behavior. It is important to note that the strength of the momentum dissipation cannot induce a MIT for given $\mathcal{K}_{1}$. This pattern bears strong resemblance to that observed in the holographic axions model with non-linear Maxwell field \cite{Baggioli:2016oju,Wu:2018zdc} or the holographic Horndeski theory with axions \cite{Jiang:2017imk,Wang:2019jyw,Zhang:2022hxl}. However, it differs from the case of the holographic EMAW (Einstein-Maxwell-axion-Weyl) theory \cite{Ling:2016dck,Wu:2018pig}, where the strength of momentum dissipation can induce a MIT for given Weyl coupling parameter.

In the right plot of Fig.~\ref{DCcalculate}, we present the DC conductivity as a function of the momentum dissipation strength $\alpha$ at zero temperature for various $\mathcal{K}_{1}$ values. Notably, it is evident that, for a given $\mathcal{K}_{1}$, the DC conductivity decreases as $\alpha$ increases. Furthermore, an interesting observation is that in the limit of large momentum dissipation, specifically at $\mathcal{K}_{1}=-1/6$, the DC conductivity tends to zero.
\begin{figure}
\center{
\includegraphics[scale=0.58]{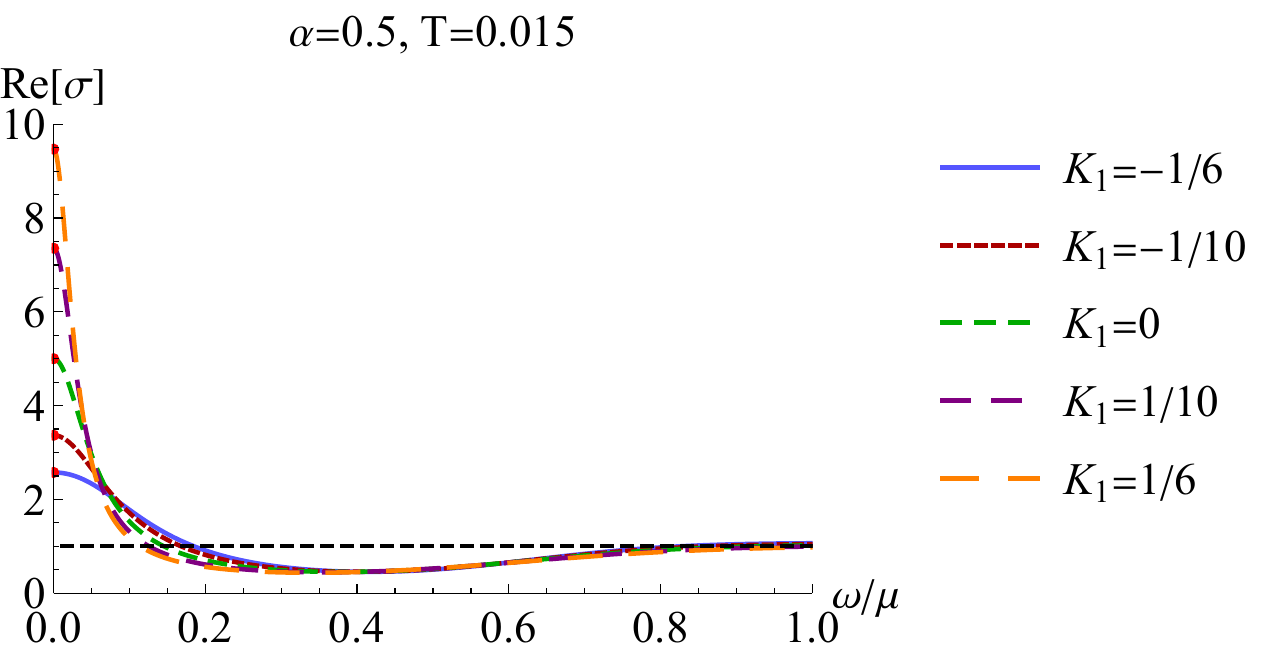}\ \hspace{0.8cm}
\includegraphics[scale=0.58]{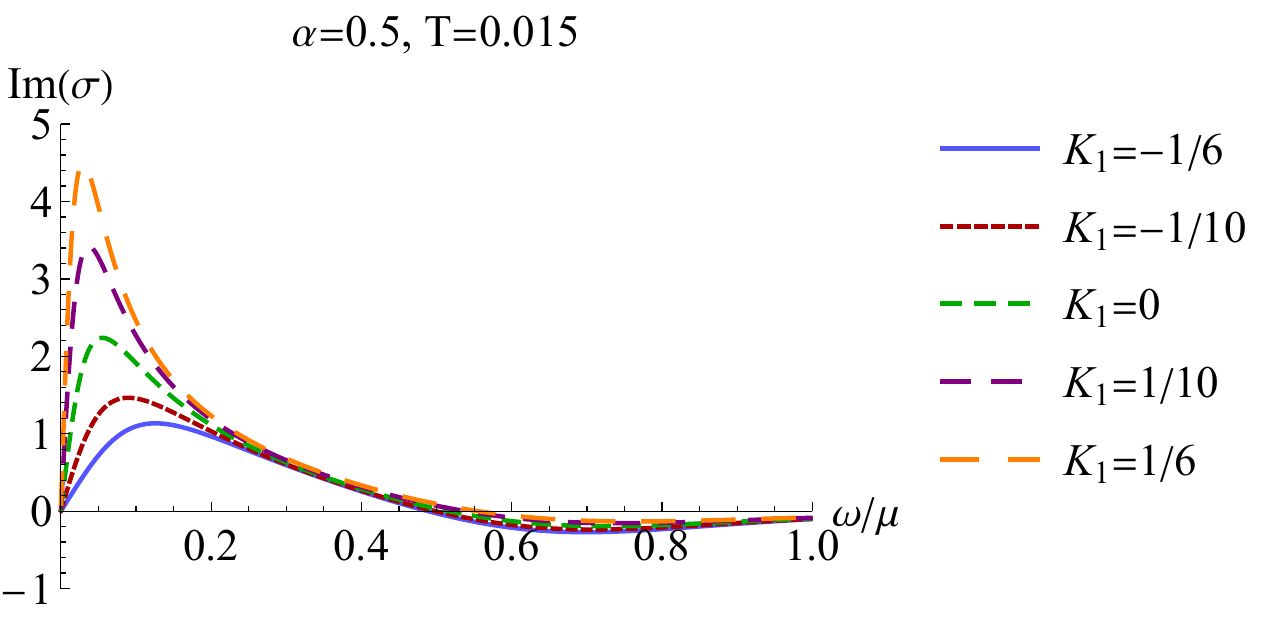}\ \\
\includegraphics[scale=0.58]{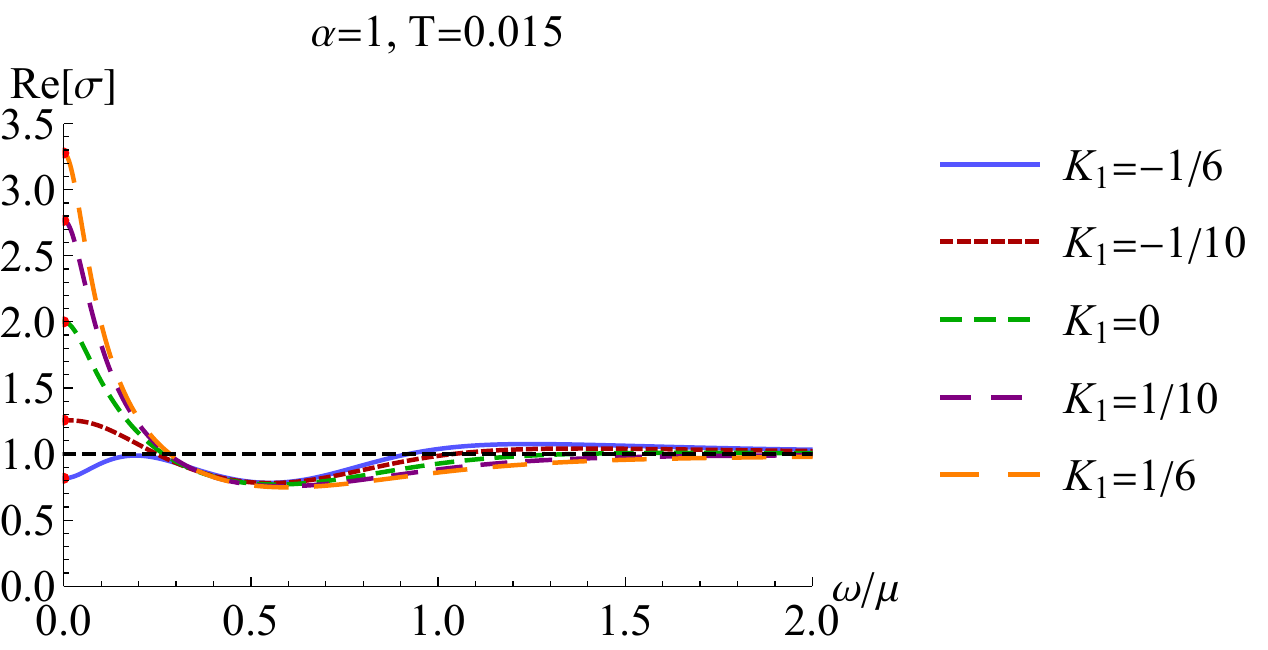}\ \hspace{0.8cm}
\includegraphics[scale=0.58]{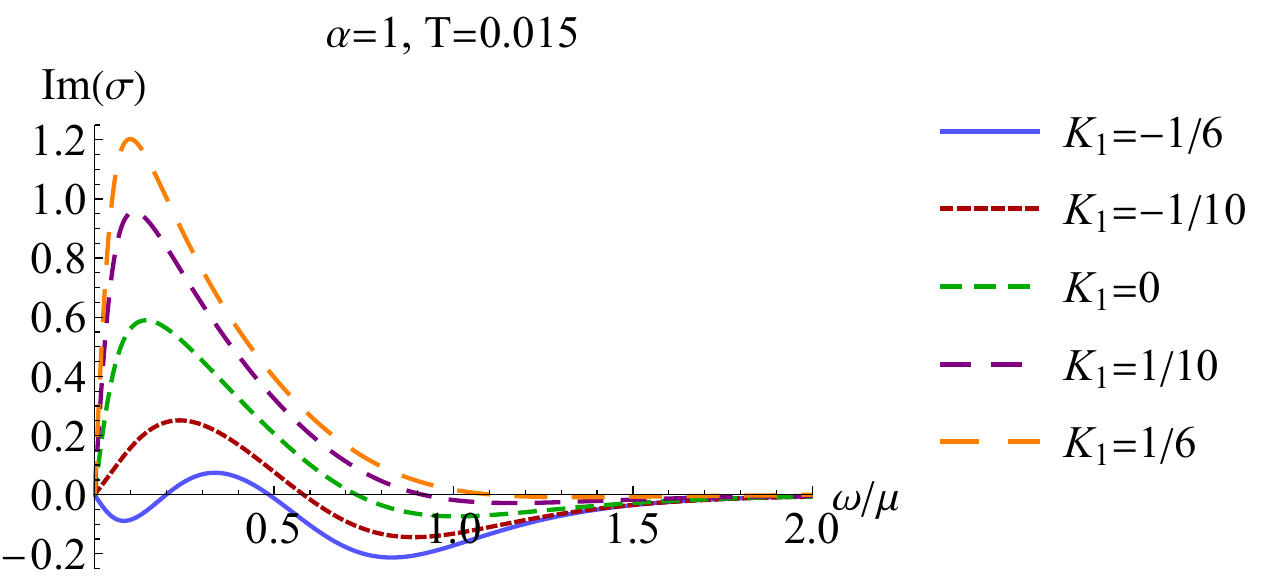}\ \\
\includegraphics[scale=0.58]{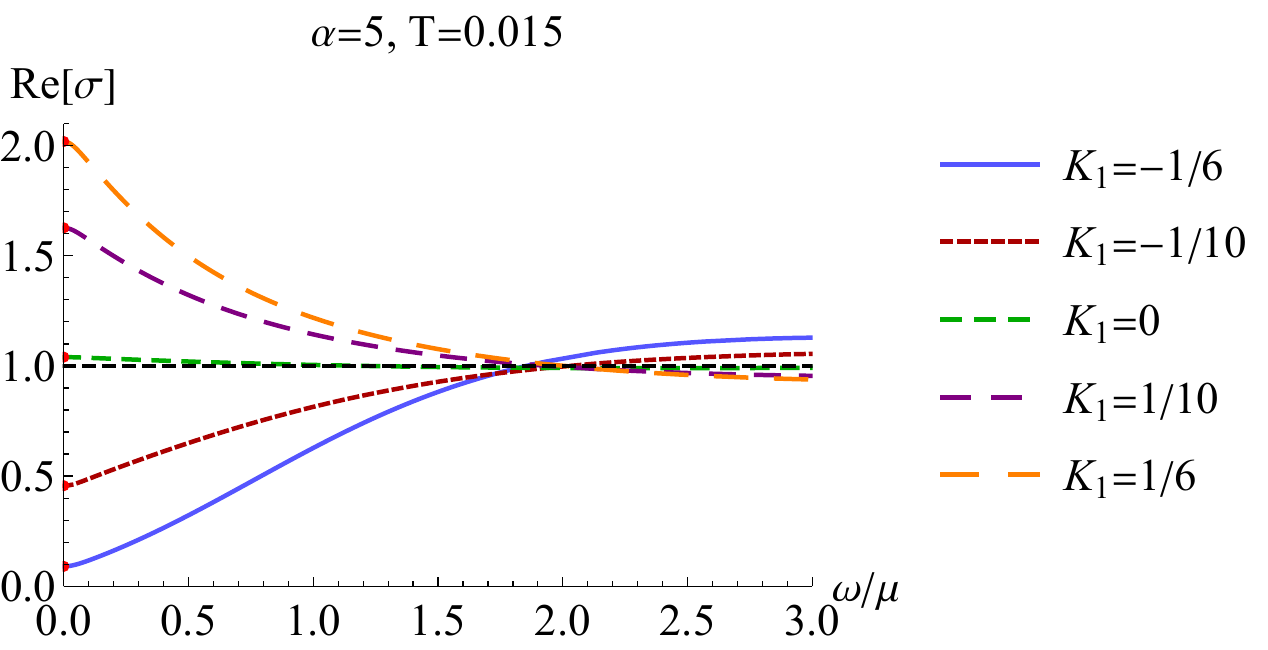}\ \hspace{0.8cm}
\includegraphics[scale=0.58]{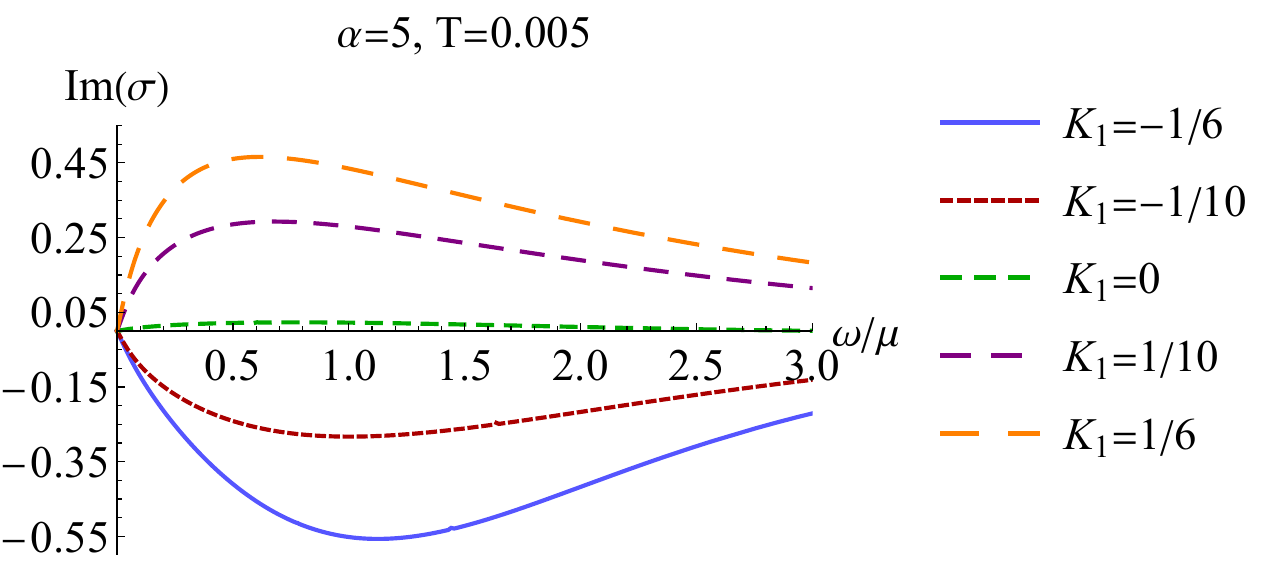}\ \\
\caption{\label{fig-AC-T0p015} The AC conductivity behavior in the normal state at $T=0.015$ with various $\mathcal{K}_{1}$ and $\alpha$. The red dots represent the corresponding values of the DC conductivity calculated using formula (\ref{DCconductivity}).}}
\end{figure}
\subsection{AC conductivity}\label{ACsection}
To compute the AC conductivity, the following forms of the time-dependent linearized perturbations would be turned on
\fa
&&
\delta A_x(t,u,x^i)=\int^{+\infty}_{-\infty}\frac{d\omega}{(2\pi)^3}e^{-i \omega t}\delta a_x(u),\nonumber\\
&&
\delta g_{tx}(t,u,x^i)=\int ^{+\infty}_{-\infty}\frac{d\omega }{(2\pi)^3}e^{-i \omega t}u^{-2}h_{tx}(u),\nonumber\\
&&
\delta X^x(t,u,x^i)=\int ^{+\infty}_{-\infty}\frac{d\omega}{(2\pi)^3}e^{-i \omega t}\delta\phi^x (u).
\ffa
This implies that there are three linearized equations of motion: one for the metric field $h_{tx}$, another for the gauge field $\delta a_x$, and the third for the scalar field $\delta\phi^x$. The asymptotic behavior of the Maxwell field near the boundary $(u\rightarrow0)$ can be expressed in the following form:
\fa
\label{ACformula}
&&
\delta a_x = a^{(0)}+a^{(1)}u+\cdot\cdot\cdot.
\ffa
According to holographic dictionary, the conductivity can be expressed as:
\fa
\sigma(\omega)=-\frac{ia^{(1)}}{\omega a^{(0)}}.
\label{sigma}
\ffa
Subsequently, we can solve the perturbative equations numerically, employing ingoing boundary conditions at the horizon. By doing so, we can extract the conductivity from Eq. \eqref{sigma}.

Figure~\ref{fig-AC-T0p015} illustrates the behaviors of the real and imaginary parts of the AC conductivity as $\alpha$ and $\mathcal{K}_{1}$ change at a temperature of $T=0.015$. In the high-frequency regime, the real parts approach unity, while the imaginary parts tend towards zero, representing universal behavior associated with the UV fixed point. However, in the low-frequency region, we observe intriguing and novel characteristics, which will be the primary focus of our subsequent analysis.

When the gauge-axion coupling $\mathcal{K}_{1}$ is turned off, i.e., $\mathcal{K}_{1}=0$, our model simplifies to the simple holographic axions model \cite{Andrade:2013gsa}. From Fig. \ref{fig-AC-T0p015} (green dashed curves representing $\mathcal{K}_{1}=0$), it is evident that a typical Drude peak emerges at low frequencies for small $\alpha$. However, as $\alpha$ increases, this Drude peak gradually diminishes and nearly vanishes for large $\alpha$. This behavior indicates that the strength of momentum dissipation induces a transition from a coherent metallic phase to an incoherent one. This observation aligns with expectations since, for small momentum dissipation, the total momentum of the system is approximately conserved, giving rise to the characteristic Drude behavior \cite{Hartnoll:2007ih,Hartnoll:2012rj,Lucas:2015pxa,Mahajan:2013cja}. With the increase in the strength of momentum dissipation, the approximate conservation of the total momentum is disrupted, leading to the breakdown of the Drude behavior \cite{Davison:2015bea,Zhou:2015qui}. These characteristics have been convincingly demonstrated in previous studies \cite{Andrade:2013gsa,Wu:2018zdc,Davison:2015bea,Zhou:2015qui,Liu:2022bam}.

When we take into account the gauge-axion coupling, the AC conductivity exhibits intriguing behaviors. 
Recalling that in our model, the strength of momentum dissipation $\alpha$ governs both the ESB and SSB, while $\mathcal{K}_{1}$ solely controls the SSB. In light of this observation, we note that the role of the coupling $\mathcal{K}_{1}$ is analogous to that of the coupling $\alpha_1$ in \cite{Liu:2022bam}. Figure~\ref{fig-AC-T0p015} depicts the frequency-dependent behavior of the AC conductivity for various combinations of $\mathcal{K}_{1}$ and $\alpha$. Notably, a distinct peak emerges in the mid-infrared (mid-IR) range for certain values of $\alpha$ and $\mathcal{K}_{1}$ (as observed in the blue curves of the second row in Fig.\ref{fig-AC-T0p015}). This mid-IR peak is attributed to the interplay between SSB and ESB, giving rise to the emergence of a pseudo-Goldstone mode\cite{Li:2018vrz,Zhong:2022mok,Liu:2022bam}, which contributes to the intriguing nature of this behavior. Furthermore, in the case where $\mathcal{K}_{1}$ is positive, we observe that as $\mathcal{K}_{1}$ or $\alpha$ increases, the low-frequency peak degrades, but it does not transform into a dip. However, when $\mathcal{K}_{1}$ becomes negative, a dip emerges in the low-frequency AC conductivity for large values of $\alpha$ or the absolute value of $\mathcal{K}_{1}$, which governs the SSB. Phenomenologically, this dip can be interpreted as indicative of vortex-like behavior \cite{Myers:2010pk,Sachdev:2011wg,Hartnoll:2016apf,Ritz:2008kh,Witczak-Krempa:2012qgh,Witczak-Krempa:2013xlz,Witczak-Krempa:2013nua,Witczak-Krempa:2013aea,Katz:2014rla,Wu:2018xjy,Damle:1997rxu,Chen:2017dsy}. Theoretically, such vortex-like behavior can be attributed to excitations arising from the Goldstone mode of SSB \cite{Zhong:2022mok}.

\section{The superfluid phase}\label{super}

In this section, we will investigate the characteristics of the superfluid state within our holographic effective theory. Previous studies \cite{Kim:2015dna,Andrade:2014xca,Baggioli:2015dwa,Kiritsis:2015hoa,Cai:2020nyd,Baggioli:2015zoa} have explored the impact of momentum dissipation. Therefore, our focus will primarily be on understanding how the interactions between the Maxwell field, axionic field, and complex scalar field influence the formation of condensation and the properties of the AC conductivity in the superfluid state.

\subsection{Condensation}

We begin by investigating the properties of condensation through numerically solving the equations of motion (\ref{Maxwell})--(\ref{Einstein}), using the ansatz (\ref{black hole background solution}). In this system, there are six parameters that influence the condensation: the strength of momentum dissipation $\alpha$, the gauge-axion coupling parameter $\mathcal{K}_1$ and $\mathcal{K}_2$, gauge parameter $\beta$, coupling parameter $n$ and the charge $q$. The effects of the gauge coupling $\beta$, coupling parameter $n$, and charge $q$ have been extensively studied in previous works (e.g., \cite{Ling:2014laa,Horowitz:2013jaa,Liu:2022bam}). For the sake of simplicity, we set $\beta=0$, $n=1$, and $q=2$ in this study. Consequently, we will systematically explore the effects of the remaining three parameters on the condensation.

\begin{figure}
  \center{
        \includegraphics[scale=0.43]{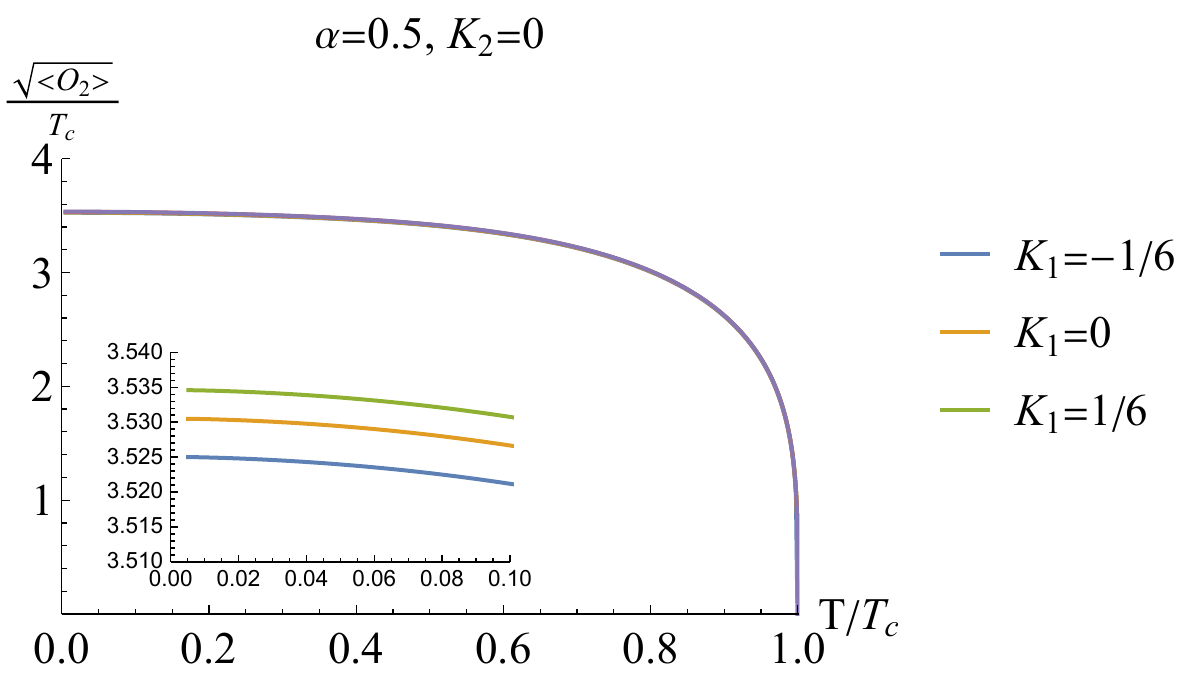}\hspace{0.1cm}
        \includegraphics[scale=0.43]{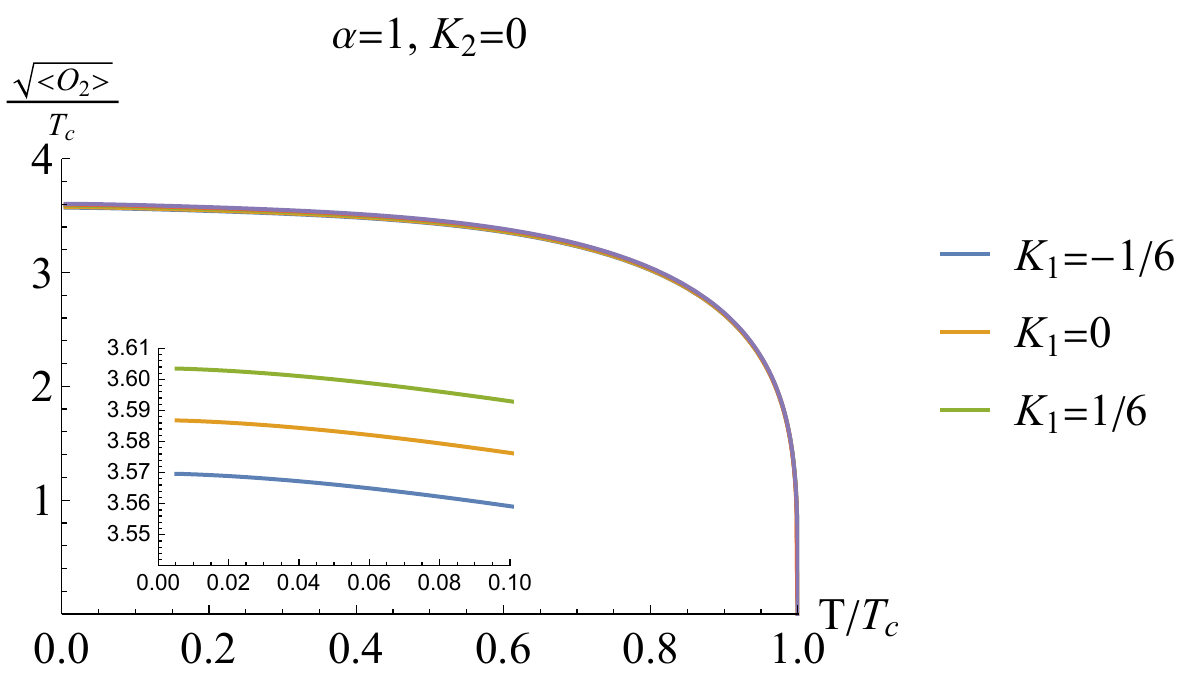}\hspace{0.1cm}
        \includegraphics[scale=0.43]{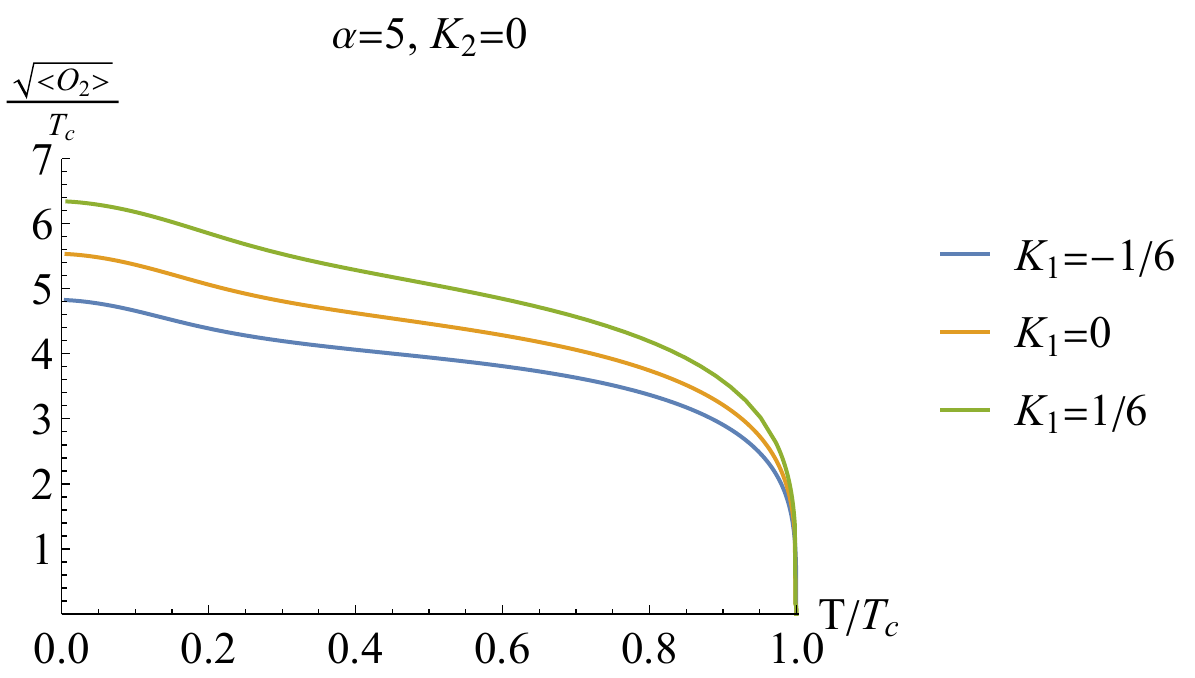}\ \\
        \caption{\label{kk1effect} The condensation $<O_2>$ as a function of temperature with varying values of $\mathcal{K}_1$ for given $\alpha$ (from left to right $\alpha=0.5$, $\alpha=1$ and $\alpha=5$). Here we set $\mathcal{K}_2=0$.}}
\end{figure}

\begin{figure}
    \center{
        \includegraphics[scale=0.43]{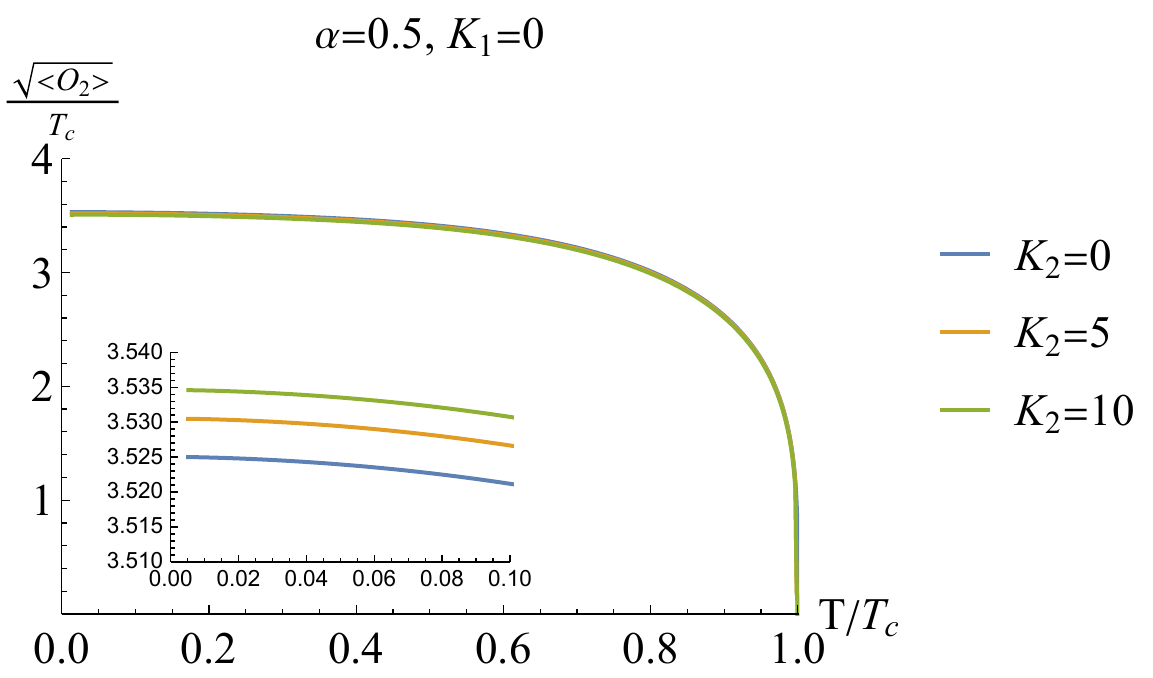}\hspace{0.1cm}
        \includegraphics[scale=0.43]{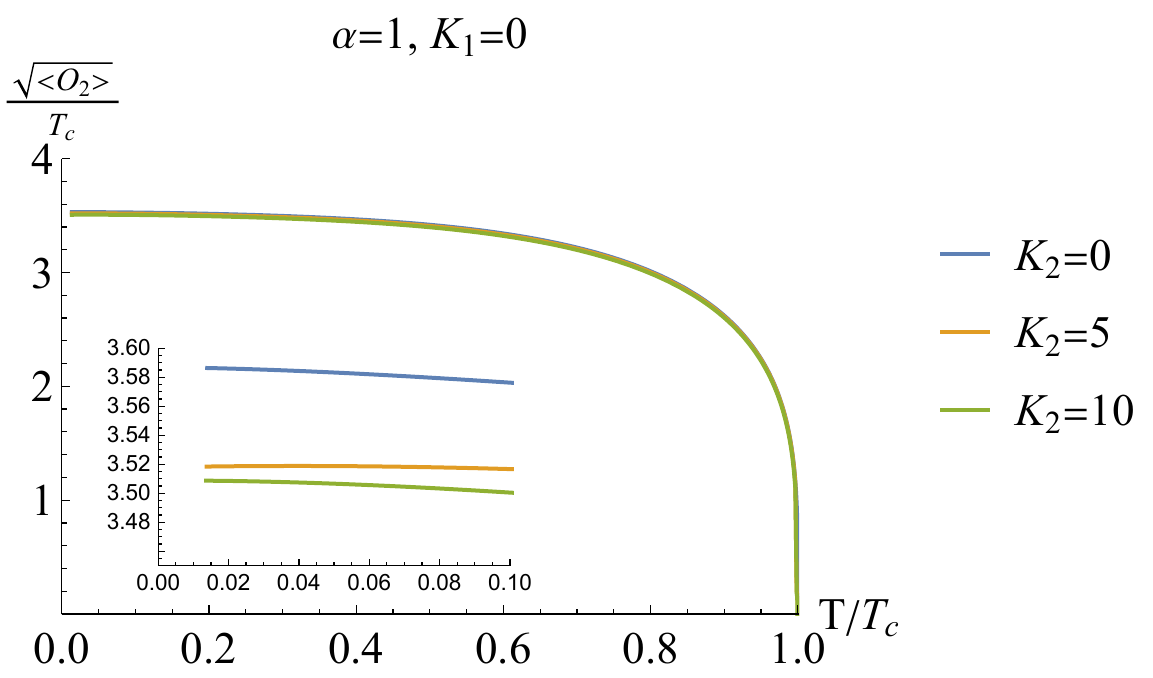}\hspace{0.1cm}
        \includegraphics[scale=0.43]{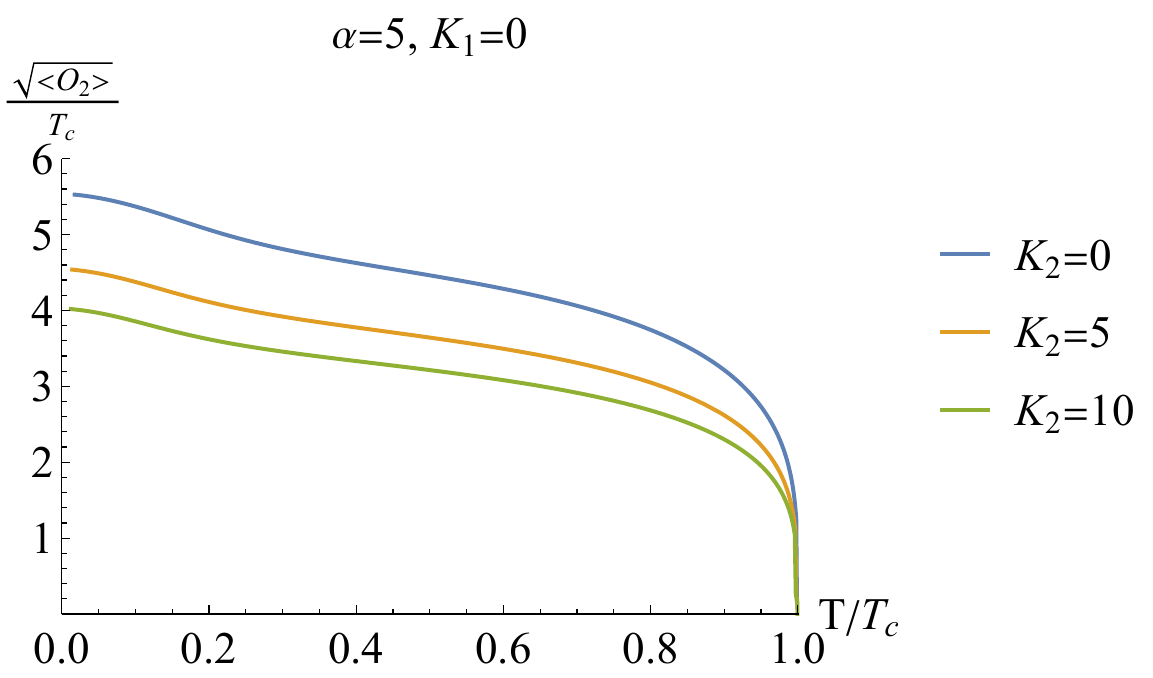}\ \\
        \caption{\label{kk2effect} The condensation $<O_2>$ as a function of temperature with varying values of $\mathcal{K}_2$ for given $\alpha$ (from left to right $\alpha=0.5$, $\alpha=1$ and $\alpha=5$). Here we set $\mathcal{K}_1=0$.}}
\end{figure}

\begin{figure}
    \center{
        \includegraphics[scale=0.69]{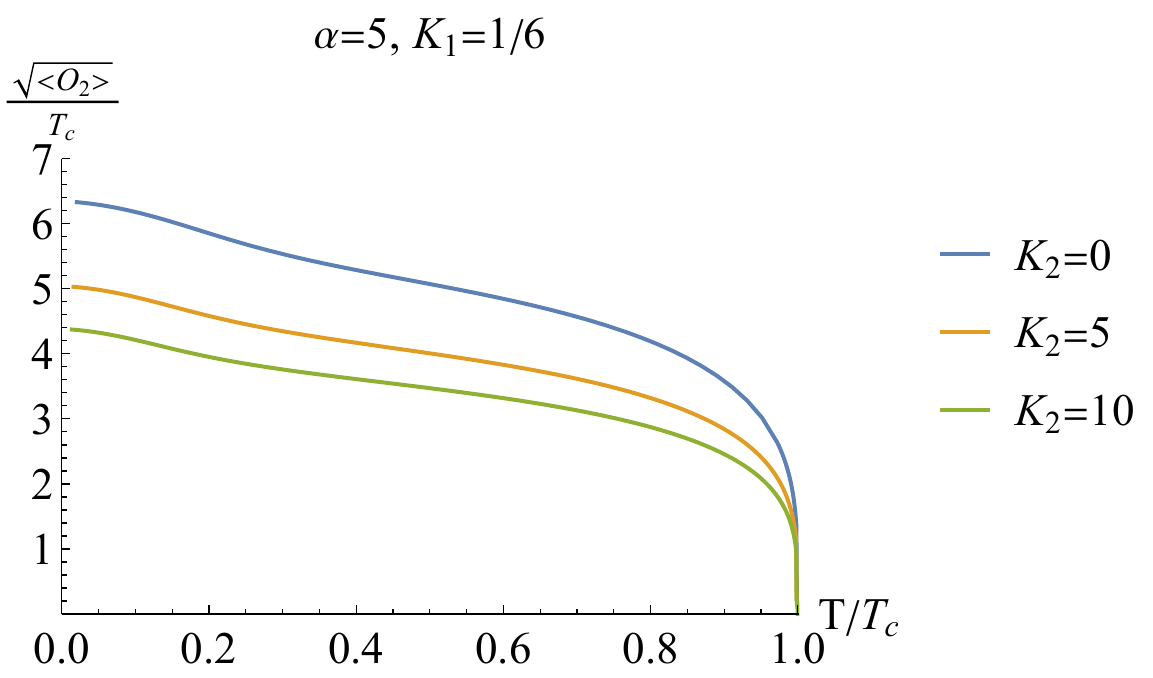}\hspace{0.1cm}
        \includegraphics[scale=0.69]{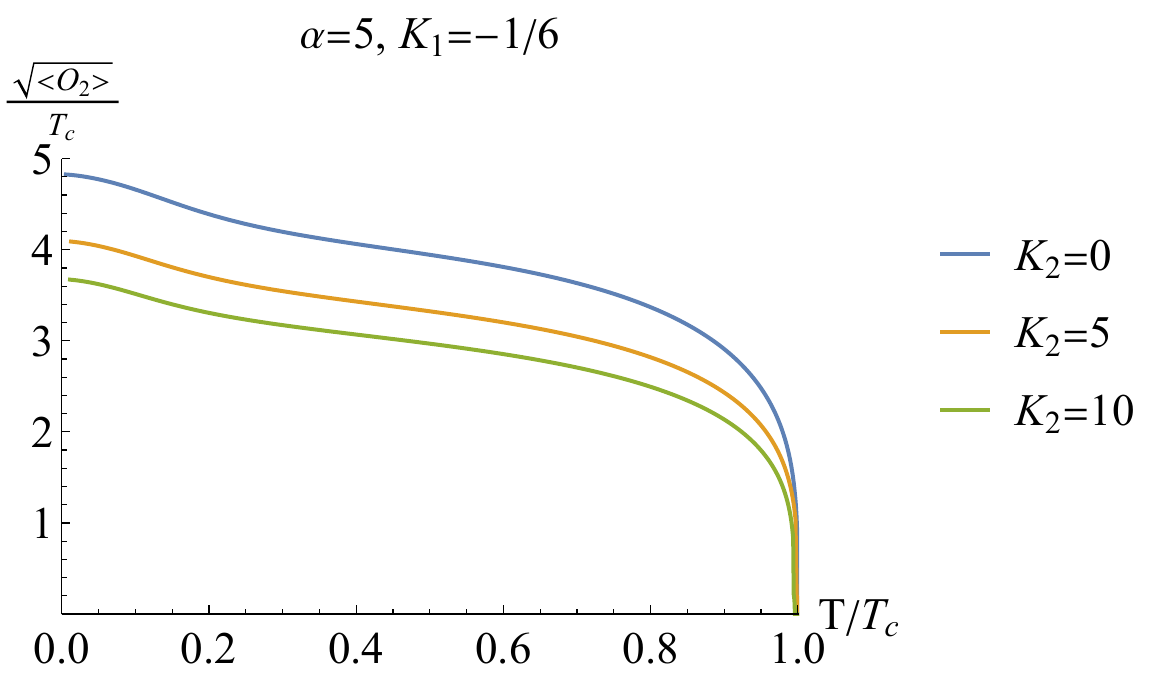}\ \\
        \caption{\label{kk1vskk2} The condensation $<O_2>$ as a function of temperature with varying values of $\mathcal{K}_2$ for given $\mathcal{K}_1$. Here we set $\alpha=5$.}}
\end{figure}

We illustrate the behavior of the condensation $<O_2>$ as a function of temperature, with temperature normalized to the critical temperature. The corresponding results are presented in Figs. \ref{kk1effect} to \ref{kk1vskk2}.
First, we explore the combined influence of momentum dissipation $\alpha$ and gauge-axion coupling $\mathcal{K}_1$ on the condensation while keeping the parameter $\mathcal{K}_2$ turned off. In Fig.~\ref{kk1effect}, we observe that the values of condensation decrease as $\mathcal{K}_1$ increases. For weak momentum dissipation, the change in condensation is relatively small. Even with momentum dissipation increasing to 1, the effect of $\mathcal{K}_1$ remains modest. However, as momentum dissipation becomes stronger, the influence of $\mathcal{K}_1$ becomes more pronounced. In other words, the effect of $\mathcal{K}_1$ on condensation is suppressed at weak momentum dissipation, while at strong momentum dissipation, the effect of $\mathcal{K}_1$ dominates over that of momentum dissipation.


Furthermore, we investigate the impact of the coupling $\mathcal{K}_2$ while setting $\mathcal{K}_1$ to zero at different strengths of momentum dissipation, as shown in Fig.~\ref{kk2effect}. Similar to $\mathcal{K}_1$, the effect of $\mathcal{K}_2$ exhibits a similar trend. At strong momentum dissipation, $\mathcal{K}_2$ dominates over the momentum dissipation, whereas at weak momentum dissipation, its effect is suppressed.

Additionally, we explore the combined influence of the couplings $\mathcal{K}_1$ and $\mathcal{K}_2$ in Fig.~\ref{kk1vskk2}, while setting $\alpha=5$ to isolate the effect of momentum dissipation. When fixing the values of $\mathcal{K}_1$, we observe that as $\mathcal{K}_2$ increases, the condensation values decrease. Notably, the running range of condensation exhibits a wider variation with increasing $\mathcal{K}_2$ for positive $\mathcal{K}_1$ compared to negative $\mathcal{K}_1$.

\subsection{The conductivity in the superfluid phase}

We follow the procedure illustrated in the normal state outlined in Sec.~\ref{ACsection} to calculate the conductivity in the superfluid phase. In this case, we focus on the influence of the three coupling parameters ($\alpha$, $\mathcal{K}_1$, $\mathcal{K}_2$) on the AC conductivity, analogous to the analysis of condensation in the previous subsection.

\begin{figure}
    \center{
        \includegraphics[scale=0.6]{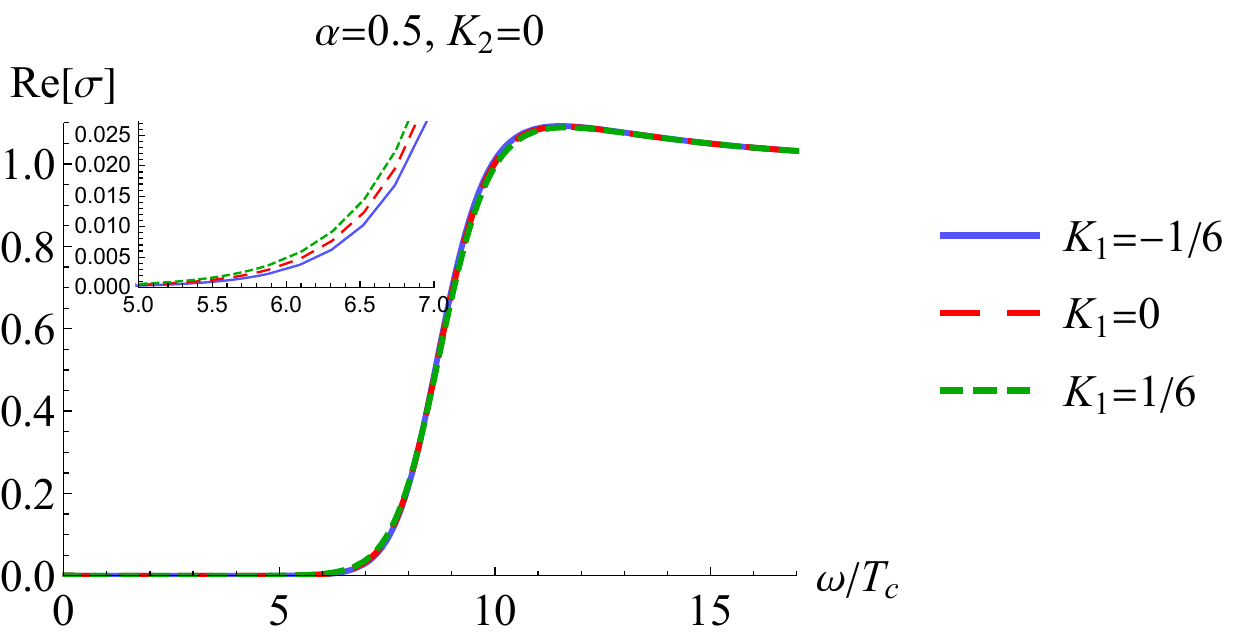}\ \hspace{0.8cm}
        \includegraphics[scale=0.6]{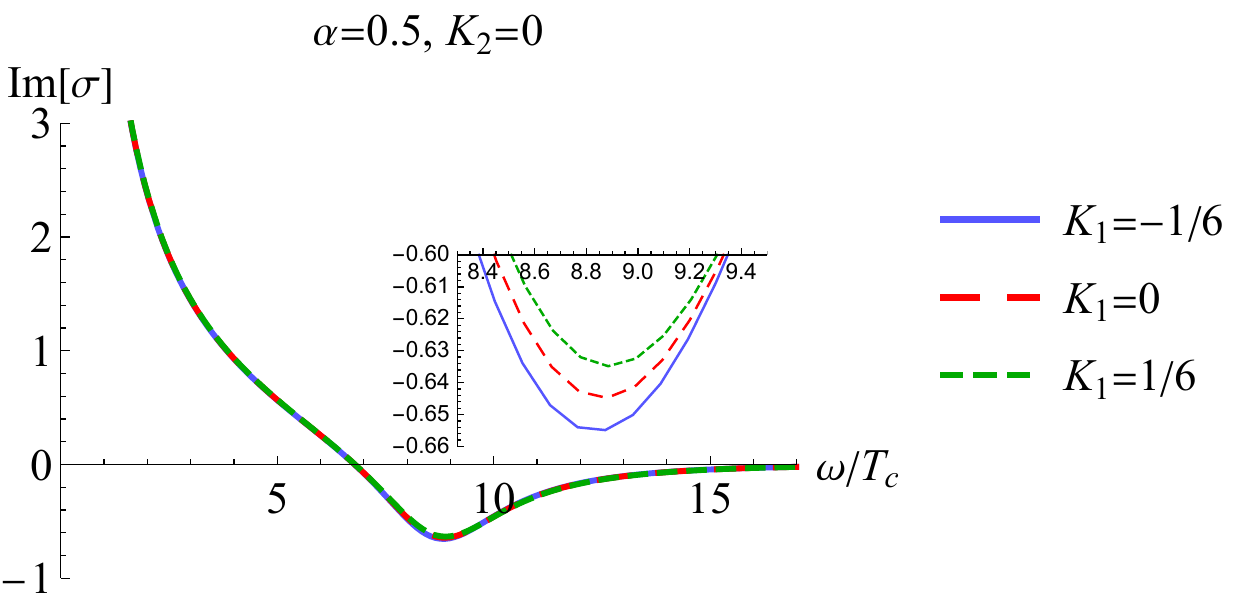}\ \\
        \includegraphics[scale=0.6]{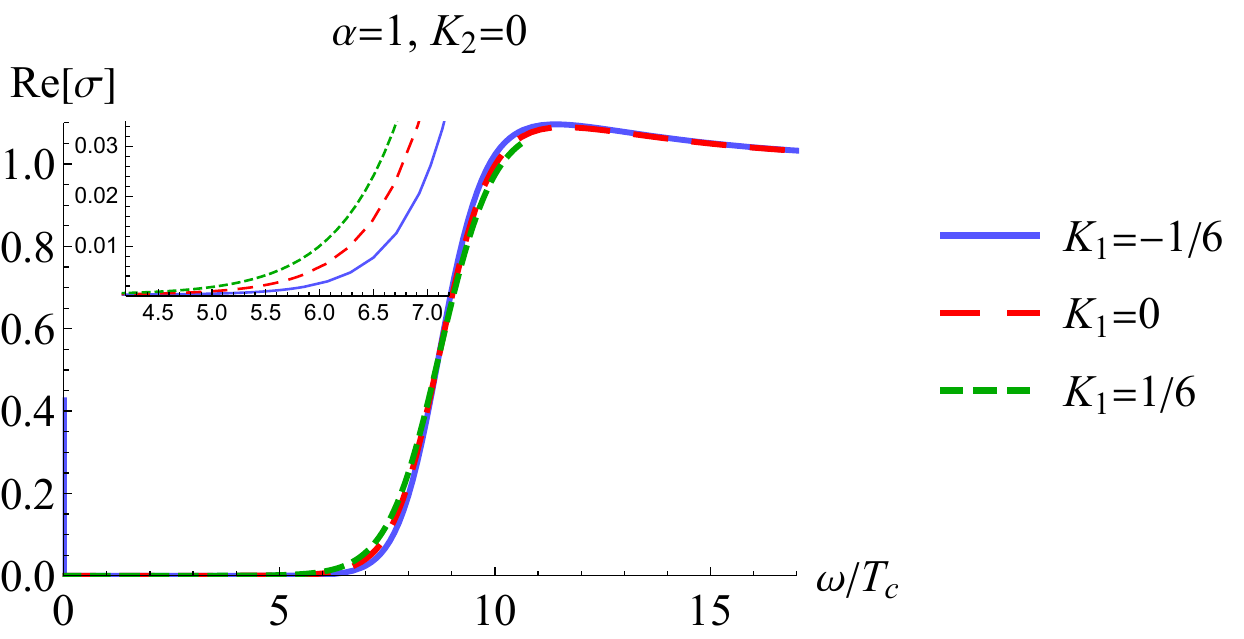}\ \hspace{0.8cm}
        \includegraphics[scale=0.6]{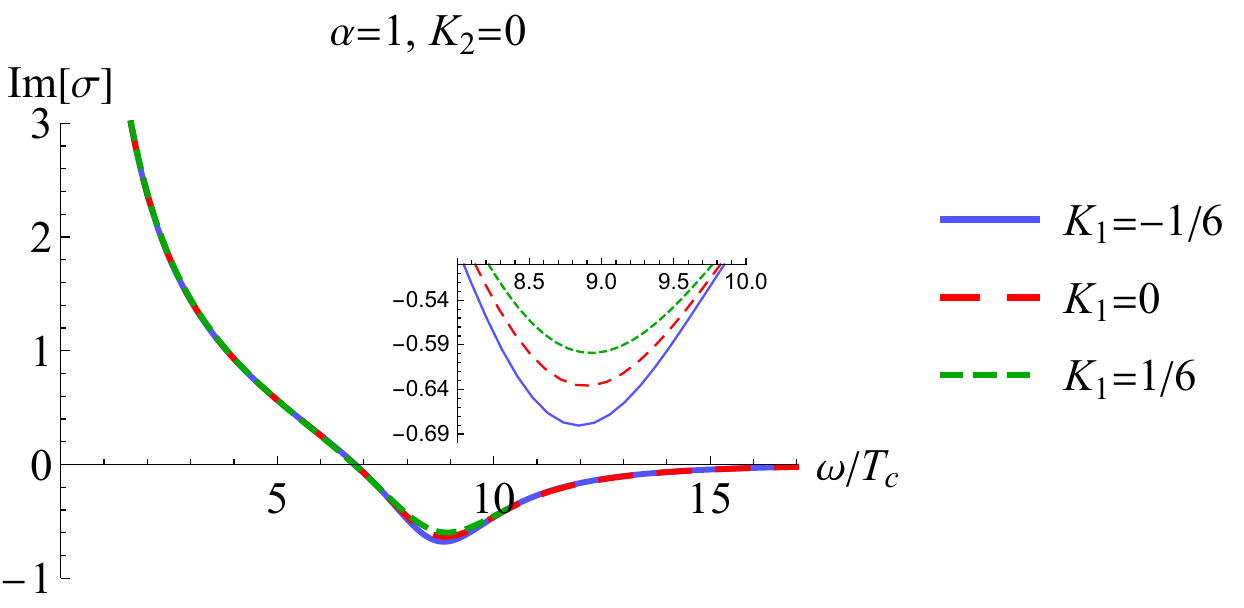}\ \\
        \includegraphics[scale=0.6]{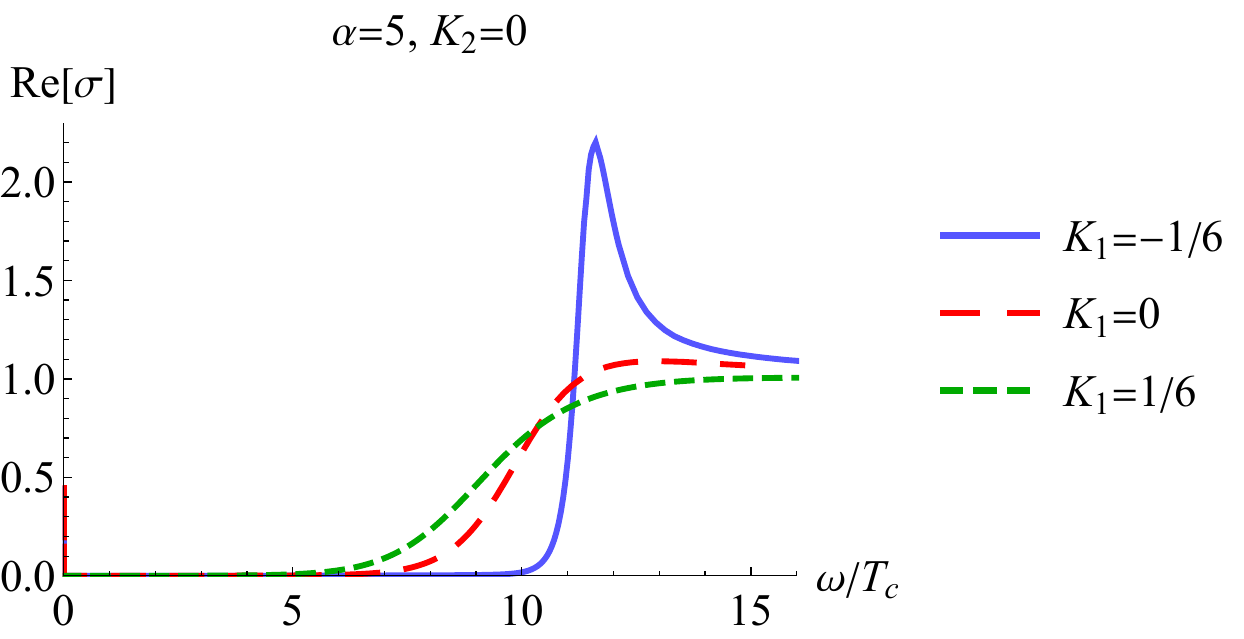}\ \hspace{0.8cm}
        \includegraphics[scale=0.6]{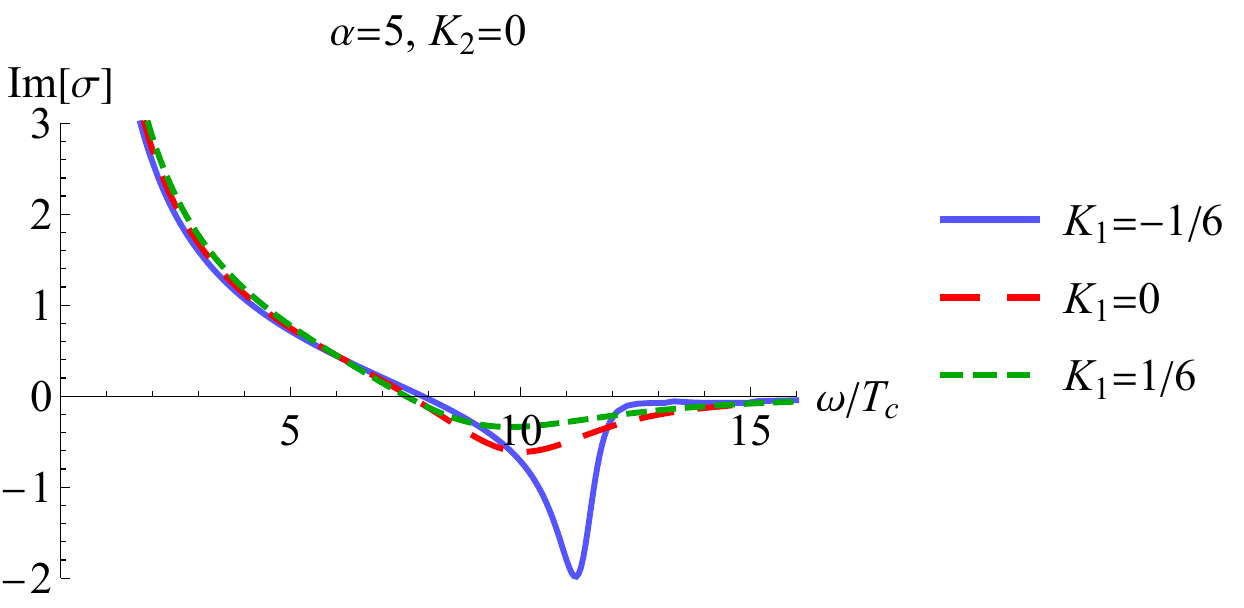}\ \\
        \caption{\label{fig-SC-kk1list} The real and imaginary parts of the conductivity, denoted as $\sigma$, are plotted as a function of frequency with different values of $\mathcal{K}_1$ for fixed $\mathcal{K}_2=0$ and varying $\alpha$. The left panels display the real part of the conductivity, while the right panels show the imaginary part of the conductivity.}}
\end{figure}
\begin{widetext}
\begin{table}[ht]
\begin{center}
\begin{tabular}{|c|c|c|c|c|}
         \hline
~$\omega_{g}/T_{c}$~ &~$\mathcal{K}_1=-1/6$&~$\mathcal{K}_1=0$~&~$\mathcal{K}_1=1/6$
          \\
        \hline
~$\alpha=0.5$~ & ~$8.8733$& ~$8.8796$~&~$8.8847$
          \\
        \hline
~$\alpha=1$~ & ~$8.8418$& ~$8.9287$~&~$8.9479$
          \\
        \hline
~$\alpha=5$~ & ~$11.2072$&~$9.9984$~&~$9.7693$
          \\
        \hline
\end{tabular}
\caption{\label{tablekk1list}The superfluid energy gap $\omega_{g}/T_{c}$ with different $\alpha$ and $\mathcal{K}_1$ at $\mathcal{K}_2=0$.
 }
\end{center}
\end{table}
\end{widetext}

First, we examine the combined impact of momentum dissipation ($\alpha$) and gauge coupling ($\mathcal{K}_1$) on the AC conductivity at a temperature of $T/T{c}\approx0.2$. Fig.~\ref{fig-SC-kk1list} displays the real and imaginary parts of the AC conductivity as functions of frequency.

When evaluating the gauge coupling $\mathcal{K}_1$ at a moderate level of momentum dissipation ($\alpha=0.5$), the presence of a gap becomes increasingly noticeable as $\mathcal{K}_1$ decreases, although the impact remains minimal. As we increase $\alpha$ to 1, the influence of $\mathcal{K}_1$ becomes stronger, yet still relatively minor. However, as momentum dissipation grows stronger, the influence of $\mathcal{K}_1$ becomes more pronounced. For negative values of $\mathcal{K}_1$, a hard-gap-like behavior at low frequencies and a prominent peak at intermediate frequencies emerge, indicating a distinct development of the superfluid component compared to positive $\mathcal{K}_1$. Notably, these features, including the hard-gap-like behavior and pronounced peak, are not observed in the $\mathcal{J}$ model explored in \cite{Liu:2022bam}. Quantitatively, under weak momentum dissipation ($\alpha=0.5$ or 1), the superfluid energy gap exhibits a slight increase as $\mathcal{K}_1$ increases. However, for strong momentum dissipation ($\alpha=5$), the superfluid energy gap significantly widens as $\mathcal{K}_1$ decreases (refer to Fig. \ref{fig-SC-kk1list} and Table \ref{tablekk1list}). Hence, we observe that the coupling $\mathcal{K}_1$ has a more substantial effect in the case of strong momentum dissipation. This finding aligns with the observations made for condensation in Fig.~\ref{kk2effect}. In summary, the effect of $\mathcal{K}_1$ depends on whether momentum dissipation is trong or weak. This effect can be ascribed to the first term in the Lagrangian density $\mathcal{L}_X$, where the interplay between $\mathcal{K}_1$ and $\alpha$ leads to a mutually reinforcing outcome. Furthermore, we particularly observe that as $\mathcal{K}_1$ approaches $-1/6$ for a fixed $\alpha$, coinciding with the point where the dual field theory becomes an insulator in the normal phase, the energy gap reaches its maximum value. This intriguingly implies that the gap, and consequently the coupling of the interactions responsible for it, become significantly stronger in the insulating state. Such a finding strongly suggests that the insulating state is a consequence of the dominance of strong-interactions. In addition, we would like to highlight that the well-defined mid-IR peak observed in the bottom-left panel of Fig. \ref{fig-SC-kk1list} lacks a known mechanism. Conducting a quasinormal modes analysis of this model could provide valuable insights into understanding these mid-IR features better. We plan to employ this analysis in our future investigations.

\begin{figure}
    \center{
        \includegraphics[scale=0.6]{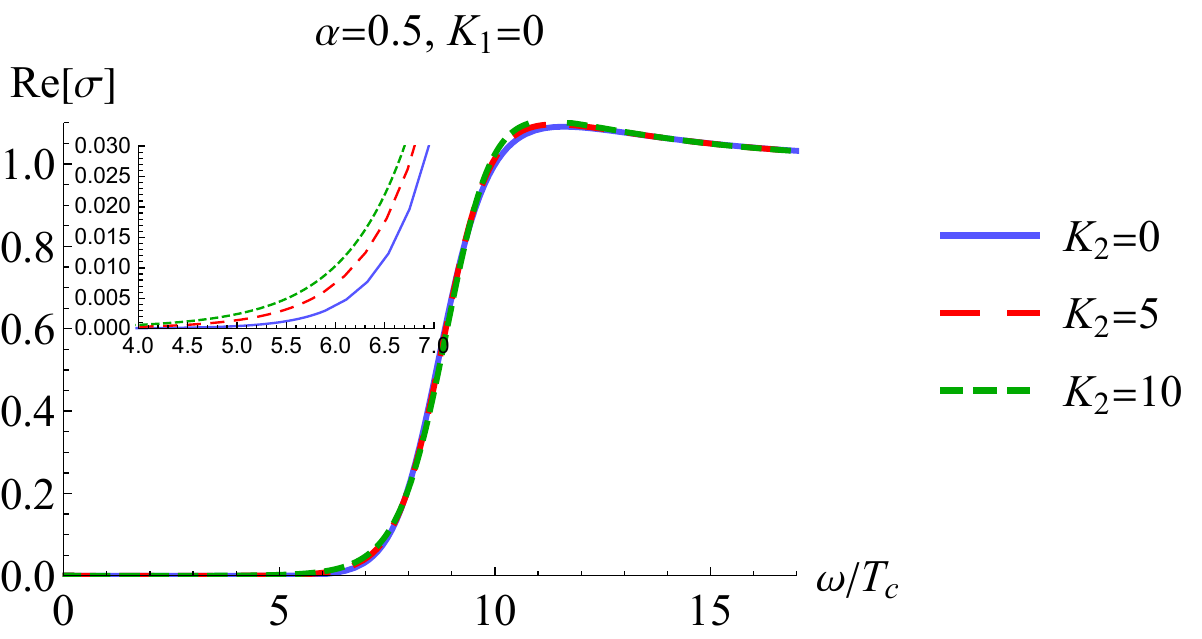}\ \hspace{0.8cm}
        \includegraphics[scale=0.6]{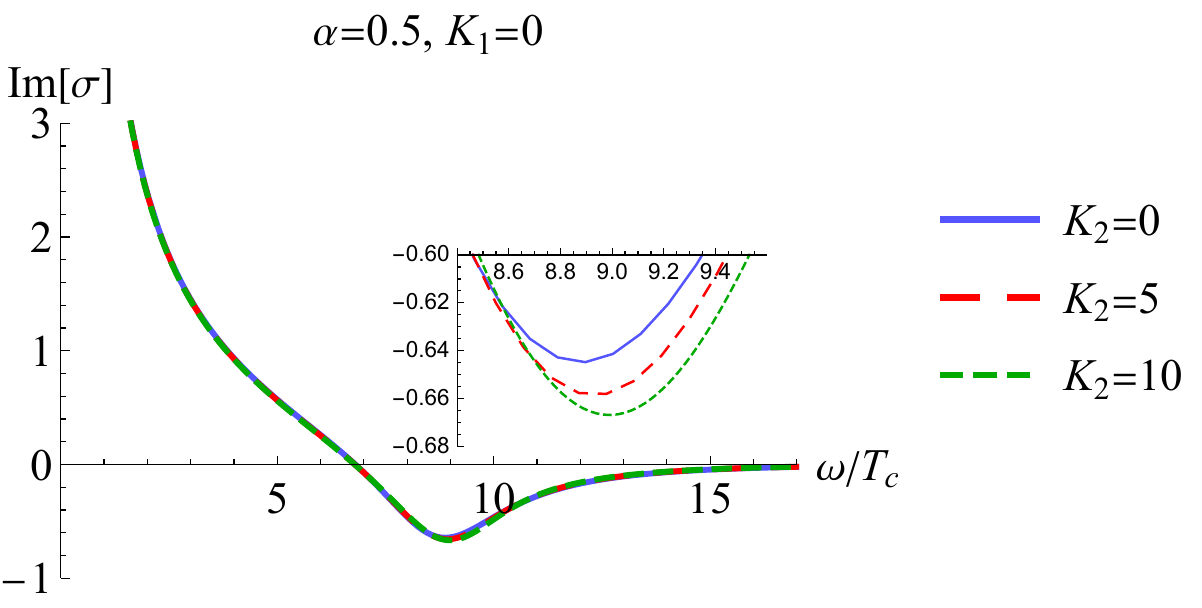}\ \\
        \includegraphics[scale=0.6]{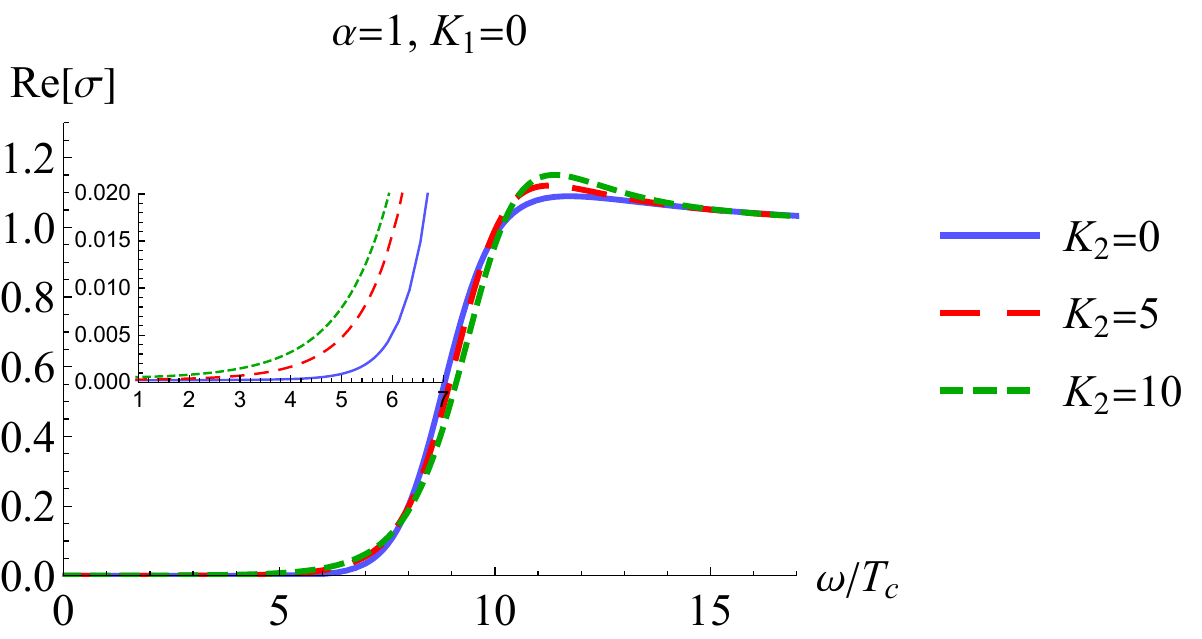}\ \hspace{0.8cm}
        \includegraphics[scale=0.6]{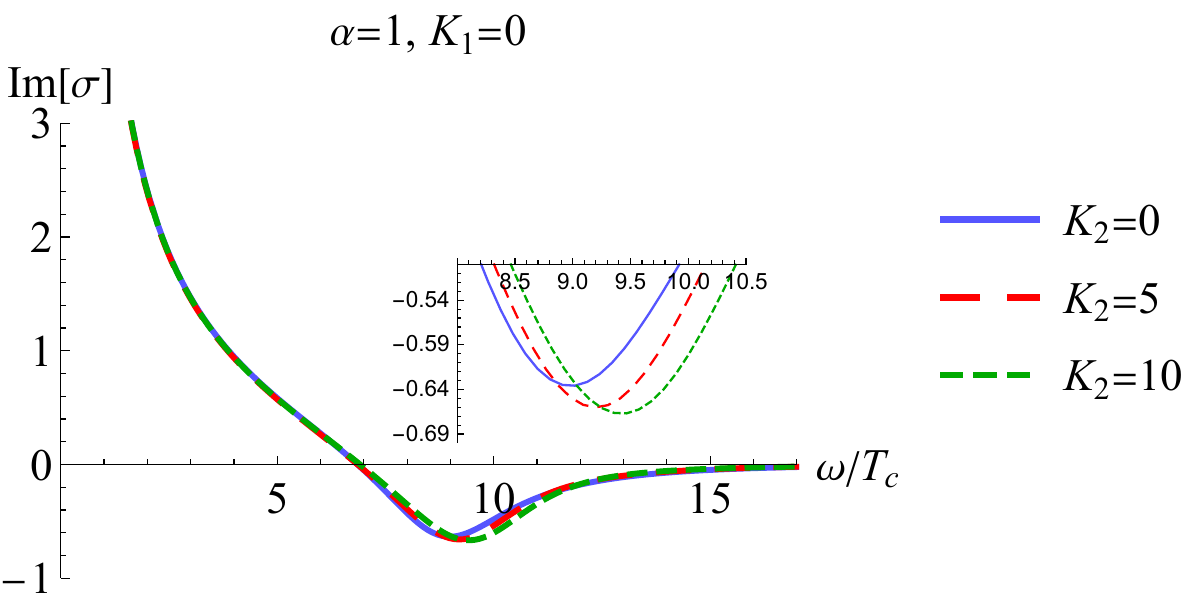}\ \\
        \includegraphics[scale=0.6]{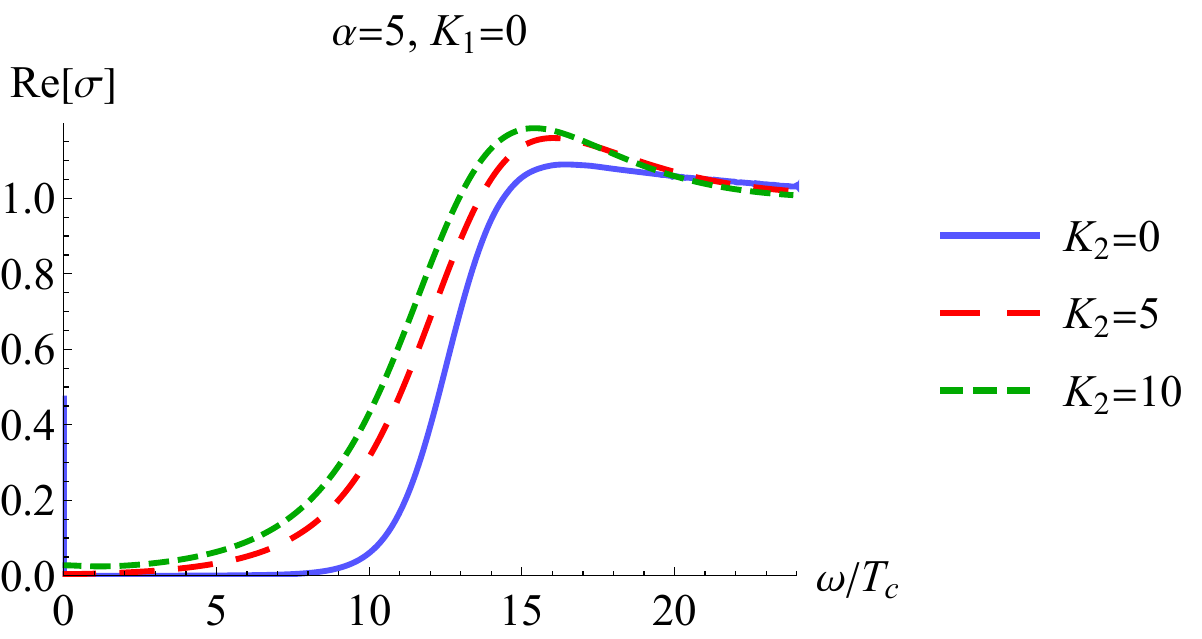}\ \hspace{0.8cm}
        \includegraphics[scale=0.6]{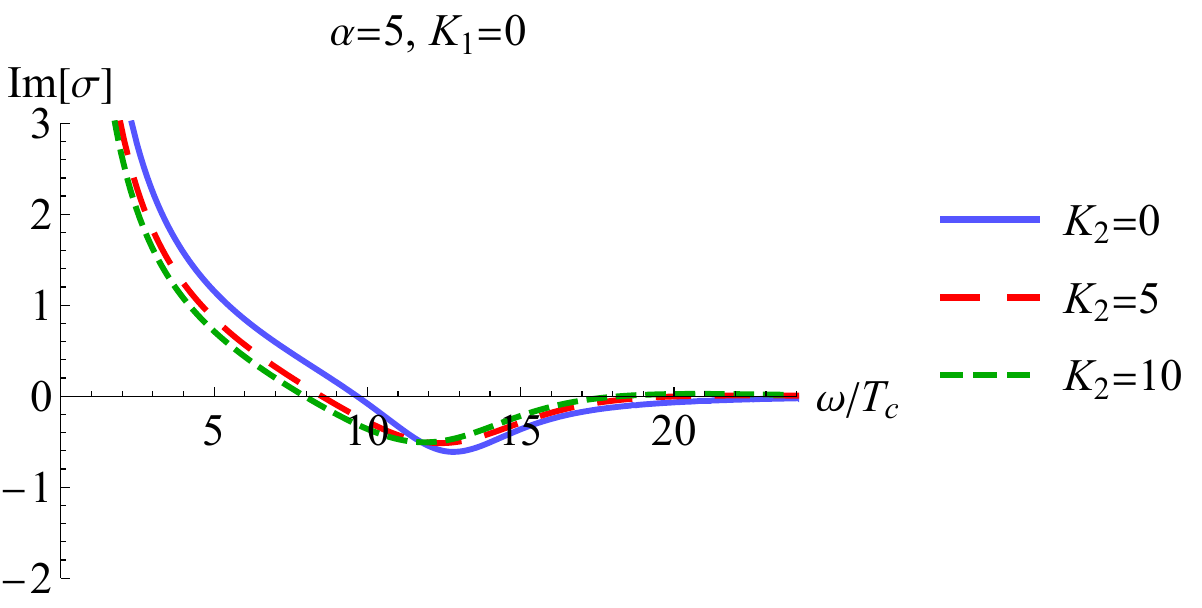}\ \\
        \caption{\label{fig-SC-nalist} The real and imaginary parts of the conductivity, denoted as $\sigma$, are plotted as a function of frequency with different values of $\mathcal{K}_2$ for fixed $\mathcal{K}_1=0$ and varying $\alpha$. The left panels display the real part of the conductivity, while the right panels show the imaginary part of the conductivity.}}
\end{figure}

Similar conclusions can be drawn when analyzing the combined effect of momentum dissipation ($\alpha$) and the coupling ($\mathcal{K}_2$) (see Fig.\ref{fig-SC-nalist} and Table\ref{tablekk2list}). However, it is worth noting that since we have constrained $\mathcal{K}_2$ to the region $\mathcal{K}_2>0$, we do not observe a pronounced peak as seen with the $\mathcal{K}_1$ coupling.

\begin{widetext}
\begin{table}[ht]
\begin{center}
\begin{tabular}{|c|c|c|c|c|}
         \hline
~$\omega_{g}/T_{c}$~ &~$\mathcal{K}_2=0$&~$\mathcal{K}_2=5$~&~$\mathcal{K}_2=10$
          \\
        \hline
~$\alpha=0.5$~ & ~$8.8959$& ~$8.9774$~&~$8.9932$
          \\
        \hline
~$\alpha=1$~ & ~$9.0191$& ~$9.1974$~&~$9.4672$
          \\
        \hline
~$\alpha=5$~ & ~$12.7968$&~$12.3933$~&~$11.8455$
          \\
        \hline
\end{tabular}
\caption{\label{tablekk2list}The superfluid energy gap $\omega_{g}/T_{c}$ with different $\alpha$ and $\mathcal{K}_2$ at $\mathcal{K}_1=0$.}
\end{center}
\end{table}
\end{widetext}

\section{Conclusions and discussions}\label{conclusion}

We have extended the investigation of an effective holographic superfluid model with gauge-axion coupling by incorporating generalized $\mathcal{K}$ coupling. The system exhibits metallic or insulating behavior in the normal state, depending on whether the coupling parameter $\mathcal{K}_1$ is positive or negative. Notably, in the limit of high momentum dissipation at $\mathcal{K}_1=-1/6$, the DC conductivity tends towards zero, displaying ideal insulating behavior.

Then, we analyze the characteristics of the AC conductivity in the normal state, where the introduction of gauge-axion coupling unveils several intriguing phenomena. For certain values of $\alpha$ and $\mathcal{K}_1$, similar to the $\mathcal{J}$ model studied in \cite{Liu:2022bam}, an intriguing mid-IR peak emerges. The formation of this mid-IR peak can be attributed to the interplay between SSB and ESB, giving rise to a pseudo-Goldstone mode. Furthermore, in the case of positive $\mathcal{K}_{1}$, we observe that as either $\mathcal{K}_{1}$ or $\alpha$ increases, the low-frequency peak diminishes without forming a dip. However, when $\mathcal{K}_{1}$ becomes negative, a dip appears in the low-frequency AC conductivity if either $\alpha$ or the absolute value of $\mathcal{K}_{1}$, which governs the SSB, is large. This dip can be phenomenologically interpreted as a vortex-like phenomenon, previously noted in \cite{Myers:2010pk,Sachdev:2011wg,Hartnoll:2016apf,Ritz:2008kh,Witczak-Krempa:2012qgh,Witczak-Krempa:2013xlz,Witczak-Krempa:2013nua,Witczak-Krempa:2013aea,Katz:2014rla,Wu:2018xjy,Damle:1997rxu,Chen:2017dsy}. Theoretically, such vortex-like patterns can be attributed to excitations arising from the Goldstone mode of SSB \cite{Zhong:2022mok}.

We further explore the influence of generalized gauge couplings on superfluid behavior. In the case of moderate momentum dissipation, the condensation only changes slightly in response to the gauge couplings.  However, when momentum dissipation becomes stronger, the impact of the gauge couplings becomes significantly more pronounced. As a result, we observe that the gauge couplings exert a more substantial influence under conditions of heightened momentum dissipation.

The AC conductivity in the superfluid phase is also investigated. With mild momentum dissipation, the superfluid energy gap experiences a slight expansion as the gauge couplings increase. However, under strong momentum dissipation, the gap widens significantly as the gauge couplings decrease. This observation aligns well with the phenomenon of condensation. It is important to emphasize that in the case of negative $\mathcal{K}_1$, distinct characteristics emerge in the form of a hard-gap-like behavior at low frequencies and a pronounced peak at intermediate frequencies. This suggests that the evolution of the superfluid component for negative $\mathcal{K}_1$ differs from that of positive $\mathcal{K}_1$. However, due to our constraint on $\mathcal{K}_2$ within the region of $\mathcal{K}_2>0$, we are unable to replicate the pronounced peak observed in the $\mathcal{K}_1$ coupling.

The coexistence and competition of ESB and SSB are the common characteristics shared by the $\mathcal{J}$ and $\mathcal{K}$ couplings, resulting in comparable phenomena in the normal state, such as a mid-IR peak and a dip in low-frequency AC conductivity. In the superfluid phase, the energy gap running with the gauge-axion coupling is another common feature shared by the $\mathcal{J}$ and $\mathcal{K}$ couplings. However, we have discovered a distinct behavior in the $\mathcal{K}$ coupling model during the superfluid phase. Specifically, a pronounced had-gap-like behavior at low frequency is observed, which differs from that observed in the $\mathcal{J}$ coupling model. This disparity can be attributed to the differences in their coupling mechanisms. A natural next step would be to extend our investigation to the system with gauge-axion coupling at finite momentum, which would allow us to further investigate the difference between the $\mathcal{J}$ and $\mathcal{K}$ couplings. We will come back this topic in the future studies.

\acknowledgments

This work is supported by the Natural Science Foundation of China (Grants Nos. 12147209 and 11975072) and the Postgraduate Research \& Practice Innovation Program of Jiangsu Province (Grant No. KYCX21\_3192). J.-P.W. is also supported by Top Talent Support Program from Yangzhou University. X.Z. is also supported by the National SKA Program of China (Grant No. 2022SKA0110203), the science research grants from the China Manned Space Project (Grant No. CMS-CSST-2021-B01), the Liaoning Revitalization Talents Program (Grant No. XLYC1905011), the National Program for Support of Top-Notch Young Professionals (Grant No. W02070050), and the National 111 Project of China (Grant No. B16009).

\bibliographystyle{style1}
\bibliography{main}
\end{document}